\documentclass[twocolumn]{aastex631}

\usepackage{lipsum}
\begin{document}

\title{Limit on Supernova Emission in the Brightest Gamma-ray Burst, GRB~221009A}
\correspondingauthor{M. Shrestha}
\email{mshrestha1@arizona.edu}

\newcommand{\LCO}{\affiliation{Las Cumbres Observatory, 6740 Cortona Drive, Suite 102, Goleta, CA 93117-5575, USA}}
\newcommand{\UCSB}{\affiliation{Department of Physics, University of California, Santa Barbara, CA 93106-9530, USA}}
\newcommand{\KITP}{\affiliation{Kavli Institute for Theoretical Physics, University of California, Santa Barbara, CA 93106-4030, USA}}
\newcommand{\UCD}{\affiliation{Department of Physics and Astronomy, University of California, Davis, 1 Shields Avenue, Davis, CA 95616-5270, USA}}
\newcommand{\WIS}{\affiliation{Department of Particle Physics and Astrophysics, Weizmann Institute of Science, 76100 Rehovot, Israel}}
\newcommand{\OKC}{\affiliation{Oskar Klein Centre, Department of Astronomy, Stockholm University, Albanova University Centre, SE-106 91 Stockholm, Sweden}}
\newcommand{\OAPD}{\affiliation{INAF-Osservatorio Astronomico di Padova, Vicolo dell'Osservatorio 5, I-35122 Padova, Italy}}
\newcommand{\Caltech}{\affiliation{Cahill Center for Astronomy and Astrophysics, California Institute of Technology, Mail Code 249-17, Pasadena, CA 91125, USA}}
\newcommand{\GSFC}{\affiliation{Astrophysics Science Division, NASA Goddard Space Flight Center, Mail Code 661, Greenbelt, MD 20771, USA}}
\newcommand{\UMD}{\affiliation{Joint Space-Science Institute, University of Maryland, College Park, MD 20742, USA}}
\newcommand{\UCB}{\affiliation{Department of Astronomy, University of California, Berkeley, CA 94720-3411, USA}}
\newcommand{\TTU}{\affiliation{Department of Physics, Texas Tech University, Box 41051, Lubbock, TX 79409-1051, USA}}
\newcommand{\STScI}{\affiliation{Space Telescope Science Institute, 3700 San Martin Drive, Baltimore, MD 21218-2410, USA}}
\newcommand{\UT}{\affiliation{University of Texas at Austin, 1 University Station C1400, Austin, TX 78712-0259, USA}}
\newcommand{\IoA}{\affiliation{Institute of Astronomy, University of Cambridge, Madingley Road, Cambridge CB3 0HA, UK}}
\newcommand{\QUB}{\affiliation{Astrophysics Research Centre, School of Mathematics and Physics, Queen's University Belfast, Belfast BT7 1NN, UK}}
\newcommand{\IPAC}{\affiliation{Spitzer Science Center, California Institute of Technology, Pasadena, CA 91125, USA}}
\newcommand{\JPL}{\affiliation{Jet Propulsion Laboratory, California Institute of Technology, 4800 Oak Grove Dr, Pasadena, CA 91109, USA}}
\newcommand{\Southampton}{\affiliation{Department of Physics and Astronomy, University of Southampton, Southampton SO17 1BJ, UK}}
\newcommand{\LANL}{\affiliation{Space and Remote Sensing, MS B244, Los Alamos National Laboratory, Los Alamos, NM 87545, USA}}
\newcommand{\Tsinghua}{\affiliation{Physics Department and Tsinghua Center for Astrophysics, Tsinghua University, Beijing, 100084, People's Republic of China}}
\newcommand{\NAOC}{\affiliation{National Astronomical Observatory of China, Chinese Academy of Sciences, Beijing, 100012, People's Republic of China}}
\newcommand{\Itagaki}{\affiliation{Itagaki Astronomical Observatory, Yamagata 990-2492, Japan}}
\newcommand{\Einstein}{\altaffiliation{Einstein Fellow}}
\newcommand{\Hubble}{\altaffiliation{Hubble Fellow}}
\newcommand{\CfA}{\affiliation{Center for Astrophysics \textbar{} Harvard \& Smithsonian, 60 Garden Street, Cambridge, MA 02138-1516, USA}}
\newcommand{\UA}{\affiliation{Steward Observatory, University of Arizona, 933 North Cherry Avenue, Tucson, AZ 85721-0065, USA}}
\newcommand{\MPIA}{\affiliation{Max-Planck-Institut f\"ur Astrophysik, Karl-Schwarzschild-Stra\ss{}e 1, D-85748 Garching, Germany}}
\newcommand{\DSFP}{\altaffiliation{LSSTC Data Science Fellow}}
\newcommand{\HCO}{\affiliation{Harvard College Observatory, 60 Garden Street, Cambridge, MA 02138-1516, USA}}
\newcommand{\Carnegie}{\affiliation{Observatories of the Carnegie Institute for Science, 813 Santa Barbara Street, Pasadena, CA 91101-1232, USA}}
\newcommand{\TAU}{\affiliation{School of Physics and Astronomy, Tel Aviv University, Tel Aviv 69978, Israel}}
\newcommand{\Edinburgh}{\affiliation{Institute for Astronomy, University of Edinburgh, Royal Observatory, Blackford Hill EH9 3HJ, UK}}
\newcommand{\Birmingham}{\affiliation{Birmingham Institute for Gravitational Wave Astronomy and School of Physics and Astronomy, University of Birmingham, Birmingham B15 2TT, UK}}
\newcommand{\Bath}{\affiliation{Department of Physics, University of Bath, Claverton Down, Bath BA2 7AY, UK}}
\newcommand{\CTIO}{\affiliation{Cerro Tololo Inter-American Observatory, National Optical Astronomy Observatory, Casilla 603, La Serena, Chile}}
\newcommand{\Potsdam}{\affiliation{Institut f\"ur Physik und Astronomie, Universit\"at Potsdam, Haus 28, Karl-Liebknecht-Str. 24/25, D-14476 Potsdam-Golm, Germany}}
\newcommand{\INPE}{\affiliation{Instituto Nacional de Pesquisas Espaciais, Avenida dos Astronautas 1758, 12227-010, S\~ao Jos\'e dos Campos -- SP, Brazil}}
\newcommand{\UNC}{\affiliation{Department of Physics and Astronomy, University of North Carolina, 120 East Cameron Avenue, Chapel Hill, NC 27599, USA}}
\newcommand{\Ohio}{\affiliation{Astrophysical Institute, Department of Physics and Astronomy, 251B Clippinger Lab, Ohio University, Athens, OH 45701-2942, USA}}
\newcommand{\AAS}{\affiliation{American Astronomical Society, 1667 K~Street NW, Suite 800, Washington, DC 20006-1681, USA}}
\newcommand{\MMT}{\affiliation{MMT and Steward Observatories, University of Arizona, 933 North Cherry Avenue, Tucson, AZ 85721-0065, USA}}
\newcommand{\Geneva}{\affiliation{ISDC, Department of Astronomy, University of Geneva, Chemin d'\'Ecogia, 16 CH-1290 Versoix, Switzerland}}
\newcommand{\IUCAA}{\affiliation{Inter-University Center for Astronomy and Astrophysics, Post Bag 4, Ganeshkhind, Pune, Maharashtra 411007, India}}
\newcommand{\CMU}{\affiliation{Department of Physics, Carnegie Mellon University, 5000 Forbes Avenue, Pittsburgh, PA 15213-3815, USA}}
\newcommand{\NAOJ}{\affiliation{Division of Science, National Astronomical Observatory of Japan, 2-21-1 Osawa, Mitaka, Tokyo 181-8588, Japan}}
\newcommand{\IfA}{\affiliation{Institute for Astronomy, University of Hawai`i, 2680 Woodlawn Drive, Honolulu, HI 96822-1839, USA}}
\newcommand{\UCSC}{\affiliation{Department of Astronomy and Astrophysics, University of California, Santa Cruz, CA 95064-1077, USA}}
\newcommand{\Purdue}{\affiliation{Department of Physics and Astronomy, Purdue University, 525 Northwestern Avenue, West Lafayette, IN 47907-2036, USA}}
\newcommand{\Princeton}{\affiliation{Department of Astrophysical Sciences, Princeton University, 4 Ivy Lane, Princeton, NJ 08540-7219, USA}}
\newcommand{\Moore}{\affiliation{Gordon and Betty Moore Foundation, 1661 Page Mill Road, Palo Alto, CA 94304-1209, USA}}
\newcommand{\Durham}{\affiliation{Department of Physics, Durham University, South Road, Durham, DH1 3LE, UK}}
\newcommand{\JHU}{\affiliation{Department of Physics and Astronomy, The Johns Hopkins University, 3400 North Charles Street, Baltimore, MD 21218, USA}}
\newcommand{\Toronto}{\affiliation{David A.\ Dunlap Department of Astronomy and Astrophysics, University of Toronto,\\ 50 St.\ George Street, Toronto, Ontario, M5S 3H4 Canada}}
\newcommand{\Duke}{\affiliation{Department of Physics, Duke University, Campus Box 90305, Durham, NC 27708, USA}}
\newcommand{\NCU}{\affiliation{Graduate Institute of Astronomy, National Central University, 300 Jhongda Road, 32001 Jhongli, Taiwan}}
\newcommand{\Columbia}{\affiliation{Department of Physics and Columbia Astrophysics Laboratory, Columbia University, Pupin Hall, New York, NY 10027, USA}}
\newcommand{\Flatiron}{\affiliation{Center for Computational Astrophysics, Flatiron Institute, 162 5th Avenue, New York, NY 10010-5902, USA}}
\newcommand{\CIERA}{\affiliation{Center for Interdisciplinary Exploration and Research in Astrophysics and Department of Physics and Astronomy, \\Northwestern University, 1800 Sherman Avenue, 8th Floor, Evanston, IL 60201, USA}}
\newcommand{\GeminiNorth}{\affiliation{Gemini Observatory, 670 North A`ohoku Place, Hilo, HI 96720-2700, USA}}
\newcommand{\Keck}{\affiliation{W.~M.~Keck Observatory, 65-1120 M\=amalahoa Highway, Kamuela, HI 96743-8431, USA}}
\newcommand{\UW}{\affiliation{Department of Astronomy, University of Washington, 3910 15th Avenue NE, Seattle, WA 98195-0002, USA}}
\newcommand{\catalyst}{\altaffiliation{LSSTC Catalyst Fellow}}
\newcommand{\USask}{\affiliation{Department of Physics \& Engineering Physics, University of Saskatchewan, 116 Science Place, Saskatoon, SK S7N 5E2, Canada}}
\newcommand{\Thacher}{\affiliation{Thacher School, 5025 Thacher Road, Ojai, CA 93023-8304, USA}}
\newcommand{\Rutgers}{\affiliation{Department of Physics and Astronomy, Rutgers, the State University of New Jersey,\\136 Frelinghuysen Road, Piscataway, NJ 08854-8019, USA}}
\newcommand{\FSU}{\affiliation{Department of Physics, Florida State University, 77 Chieftan Way, Tallahassee, FL 32306-4350, USA}}
\newcommand{\Melbourne}{\affiliation{School of Physics, The University of Melbourne, Parkville, VIC 3010, Australia}}
\newcommand{\ASTROthreeD}{\affiliation{ARC Centre of Excellence for All Sky Astrophysics in 3 Dimensions (ASTRO 3D)}}
\newcommand{\Stromlo}{\affiliation{Mt.\ Stromlo Observatory, The Research School of Astronomy and Astrophysics, Australian National University, ACT 2601, Australia}}
\newcommand{\NCPAS}{\affiliation{National Centre for the Public Awareness of Science, Australian National University, Canberra, ACT 2611, Australia}}
\newcommand{\TAMU}{\affiliation{Department of Physics and Astronomy, Texas A\&M University, 4242 TAMU, College Station, TX 77843, USA}}
\newcommand{\Mitchell}{\affiliation{George P.\ and Cynthia Woods Mitchell Institute for Fundamental Physics \& Astronomy, College Station, TX 77843, USA}}
\newcommand{\ESO}{\affiliation{European Southern Observatory, Alonso de C\'ordova 3107, Casilla 19, Santiago, Chile}}
\newcommand{\ICE}{\affiliation{Institute of Space Sciences (ICE, CSIC), Campus UAB, Carrer
de Can Magrans, s/n, E-08193 Barcelona, Spain}}
\newcommand{\IEEC}{\affiliation{Institut d'Estudis Espacials de Catalunya, Gran Capit\`a, 2-4, Edifici Nexus, Desp.\ 201, E-08034 Barcelona, Spain}}
\newcommand{\Warwick}{\affiliation{Department of Physics, University of Warwick, Gibbet Hill Road, Coventry CV4 7AL, UK}}
\newcommand{\Macquarie}{\affiliation{School of Mathematical and Physical Sciences, Macquarie University, NSW 2109, Australia}}
\newcommand{\AAARC}{\affiliation{Astronomy, Astrophysics and Astrophotonics Research Centre, Macquarie University, Sydney, NSW 2109, Australia}}
\newcommand{\Capodimonte}{\affiliation{INAF - Capodimonte Astronomical Observatory, Salita Moiariello 16, I-80131 Napoli, Italy}}
\newcommand{\INFNNapoli}{\affiliation{INFN - Napoli, Strada Comunale Cinthia, I-80126 Napoli, Italy}}
\newcommand{\ICRANet}{\affiliation{ICRANet, Piazza della Repubblica 10, I-65122 Pescara, Italy}}
\newcommand{\MSU}{\affiliation{Center for Data Intensive and Time Domain Astronomy, Department of Physics and Astronomy,\\Michigan State University, East Lansing, MI 48824, USA}}
\newcommand{\SETI}{\affiliation{SETI Institute,
339 Bernardo Ave, Suite 200, Mountain View, CA 94043, USA}}
\newcommand{\IAIFI}{\affiliation{The NSF AI Institute for Artificial Intelligence and Fundamental Interactions}}
\newcommand{\ANUC}{\affiliation{Department of Astronomy, AlbaNova University Center, Stockholm University, SE-10691 Stockholm, Sweden}}

\newcommand{\Konkoly}{\affiliation{Konkoly Observatory,  CSFK, Konkoly-Thege M. \'ut 15-17, Budapest, 1121, Hungary}}
\newcommand{\ELTE}{\affiliation{ELTE E\"otv\"os Lor\'and University, Institute of Physics, P\'azm\'any P\'eter s\'et\'any 1/A, Budapest, 1117 Hungary}}
\newcommand{\SZTE}{\affiliation{Department of Experimental Physics, University of Szeged, D\'om t\'er 9, Szeged, 6720, Hungary}}
\newcommand{\IdAlta}{\affiliation{Instituto de Alta Investigaci\'on, Sede Esmeralda, Universidad de Tarapac\'a, Av. Luis Emilio Recabarren 2477, Iquique, Chile}}
\newcommand{\Kavli}{\affiliation{Kavli Institute for Cosmological Physics, University of Chicago, Chicago, IL 60637, USA}}
\newcommand{\UofChicago}{\affiliation{Department of Astronomy and Astrophysics, University of Chicago, Chicago, IL 60637, USA}}
\newcommand{\Fermi}{\affiliation{Fermi National Accelerator Laboratory, P.O.\ Box 500, Batavia, IL 60510, USA}}
\newcommand{\Dartmouth}{\affiliation{Department of Physics and Astronomy, Dartmouth College, Hanover, NH 03755, USA}}
\newcommand{\Surrey}{\affiliation{Department of Physics, University of Surrey, Guildford GU2 7XH, UK}}
\author[0000-0002-4022-1874]{Manisha Shrestha}
\UA
\author[0000-0003-4102-380X]{David J. Sand}
\UA
\author[0000-0002-8297-2473]{Kate D. Alexander}
\UA
\author[0000-0002-4924-444X]{K. Azalee Bostroem}
\catalyst\UA
\author[0000-0002-0832-2974]{Griffin Hosseinzadeh}
\UA
\author[0000-0002-0744-0047]{Jeniveve Pearson}
\UA
\author[0000-0001-8341-3940]{Mojgan Aghakhanloo}
\UA
\author[0000-0001-8764-7832]{J{\'o}zsef Vink{\'o}}
\Konkoly
\ELTE
\UT
\SZTE

\author[0000-0003-0123-0062]{Jennifer E. Andrews}
\GeminiNorth

\author[0000-0001-5754-4007]{Jacob E. Jencson}
\JHU

\author[0000-0001-9589-3793]{M.~J. Lundquist}
\Keck
\author[0000-0003-2732-4956]{Samuel Wyatt}
\UW
\author[0000-0003-4253-656X]{D.\ Andrew Howell}
\LCO\UCSB
\author[0000-0001-5807-7893]{Curtis McCully}
\LCO
\UCSB

\author[0000-0003-0209-9246]{Estefania Padilla Gonzalez}
\LCO
\UCSB
\author[0000-0002-7472-1279]{Craig Pellegrino}
\LCO 
\UCSB
\author[0000-0003-0794-5982]{Giacomo Terreran}
\LCO 
\UCSB

\author[0000-0002-1125-9187]{Daichi Hiramatsu}
\CfA
\IAIFI

\author{Megan Newsome}
\LCO 
\UCSB

\author{Joseph Farah}
\LCO 
\UCSB

\author[0000-0001-8738-6011]{Saurabh W.\ Jha}
\Rutgers

\author[0000-0001-5510-2424]{Nathan Smith}
\UA

\author[0000-0003-1349-6538]{J.\ Craig Wheeler}
\UT

\author[0000-0002-9144-7726]{Clara Mart\'{i}nez-V\'{a}zquez}
\GeminiNorth

\author[0000-0002-3690-105X]{Julio a. Carballo-Bello}
\IdAlta

\author[0000-0001-8251-933X]{Alex Drlica-Wagner}
\Fermi
\Kavli
\UofChicago

\author[0000-0001-5160-4486]{David J. James}
\affiliation{ASTRAVEO LLC, PO Box 1668, MA 01931}
\affiliation{Applied Materials Inc., 35 Dory Road, Gloucester, MA 01930}

\author[0000-0001-9649-4815]{Burçin Mutlu-Pakdil}
\Dartmouth

\author[0000-0003-1479-3059]{Guy S. Stringfellow}
\affiliation{Center for Astrophysics and Space Astronomy, University of Colorado Boulder, UCB 389,Boulder, Colorado, 80309-0389 USA}

\author[0000-0002-1594-1466]{Joanna D. Sakowska}
\Surrey

\author[0000-0002-8282-469X]{Noelia e. D. No\"{e}l}
\Surrey

\author[0000-0003-4383-2969]{Cl\'{e}cio R. Bom}
\affiliation{Centro Brasileiro de Pesquisas F\'isicas, Rua Dr. Xavier Sigaud 150, 22290-180 Rio de Janeiro, RJ, Brazil}

\author{Kyler Kuehn}
\affiliation{Lowell Observatory, 1400 W Mars Hill Rd, Flagstaff, AZ 86001}
\affiliation{Australian Astronomical Optics, Macquarie University, North Ryde, NSW 2113, Australia}








\begin{abstract}
We present photometric and spectroscopic observations of the extraordinary gamma-ray burst (GRB) 221009A in search of an associated supernova. Some past GRBs have shown bumps in the optical light curve that coincide with the emergence of supernova spectral features, but we do not detect any significant light curve features in GRB~221009A, nor do we detect any clear sign of supernova spectral features. Using two well-studied GRB-associated supernovae (SN~2013dx, $M_{r,max} = -19.54$; SN~2016jca, $M_{r,max} = -19.04$) at a similar redshift as GRB~221009A ($z=0.151$), we modeled how the emergence of a supernova would affect the light curve. If we assume the GRB afterglow to decay at the same rate as the X-ray data, the combination of afterglow and a supernova component is fainter than the observed GRB brightness. For the case where we assume the best-fit power law to the optical data as the GRB afterglow component, a supernova contribution should have created a clear bump in the light curve, assuming only extinction from the Milky Way.  If we assume a higher extinction of $E(B-V)$=$1.74$ mag (as has been suggested elsewhere), the supernova contribution would have been hard to detect, with a limit on the associated supernova  of $M_{r,max} \approx-$19.54. We do not observe any clear supernova features in our spectra, which were taken around the time of expected maximum light. The lack of a bright supernova associated with GRB~221009A may indicate that the energy from the explosion is mostly concentrated in the jet, leaving a lower energy budget available for the supernova. 
\end{abstract}



\section{Introduction} \label{sec:intro}
Long gamma-ray bursts (GRBs) are thought to be produced by the explosion of very massive stars \citep{Woosley_2006,Hjorth_2012}. These explosions produce relativistic jets where internal dissipation via synchrotron radiation creates prompt emission in $\gamma$ rays on timescales of seconds. Subsequently, the interaction of the jet ejecta with the ambient medium produces afterglow emission across the electromagnetic spectrum lasting hours to weeks. A few days to weeks after the explosion, an emerging supernova (SN) is often observed as an excess above the afterglow emission.  

Long GRBs are often associated with type Ic-BL SNe. Supernovae of this type are core-collapse events where the progenitor star has lost a significant amount of its hydrogen (H) and helium (He) envelope \citep{Filippenko_1997}.  There are competing, plausible progenitor scenarios for SN Ic-BL, including single Wolf-Rayet stars \citep{Gaskell_1986,Smartt_2009} or binary massive stars \citep{Podsiadlowski_1993,Nomoto_1995}, which can explain the stripping of the H and He envelope. However, to date, the progenitors of these SNe have not been definitively identified \citep[see][for a review]{Smartt_2009}.  These supernovae have high ejecta expansion velocities of order 15,000-30,000 km s$^{-1}$ \citep{Drout_2011, Modjaz_2011,Modjaz_2016,Cano_2017} which lead to broad spectra features. Interestingly, only SNe of type Ic-BL have been observed in association with long GRBs, however, many SNe Ic-BL have been observed without a GRB component \citep{Modjaz_2020}. The fraction of long GRBs that are accompanied by a SN is still debated. \citet{Rossi_2022b} found that for over 1400 long GRBs discovered by Swift through 2022, only 40-50 of them have associated SNe identified via a bump in the optical light curve, and 28 have spectroscopic confirmation. \cite{Dado_2018} found that for low redshifts, the number of GRBs with an associated supernovae is comparable to GRBs without supernovae. 

However, where deep spectroscopic observations are possible (low $z$ cases), there are only 4 GRBs with no associated supernovae: GRB~060505 \citep{Fynbo_2006}, GRB~060614 \citep{Fynbo_2006,Gal_2006,Della_2006}, GRB~111005A \citep{Michal_2018,Tanga_2018}, and GRB~211211A \citep[][which instead showed potential kilonova emission]{Rastinejad_2022_kilonova, Troja_2022}. For GRBs without deep spectroscopic observations, the identification of a supernova bump in the lightcurve based purely on photometry can be challenging, and often depends sensitively on the assumptions made when modeling the GRB afterglow. For example, \cite{Melandri_2022} presented the results of a SN connected to the GRB190114C (the first GRB with detected TeV emission) using various facilities, including HST. They find a large range in the probable luminosity of the associated supernova SN~2019jrj, largely due to uncertainties in estimating the time of the GRB jet break. Their study shows that late-time photometry is critical for constraining the jet break time, which in turn helps to constrain the energy of the SN explosion connected to the GRB. 

The recent GRB~221009A (RA (J2000)= 19:13:03.50, Dec (J2000) = +19:46:24.23; \citealt{Laskar_2022}), at a redshift of z = 0.151 \citep{postigo_2022, Tirado_2022}, provides a unique opportunity to explore GRB physics in detail. In contrast to many other nearby GRBs, which are often underluminous compared to more distant ``cosmological" GRBs at $z>1$ \citep{Dainotti_2022}, GRB~221009A is the brightest GRB to ever be detected by the Fermi Gamma-ray Burst Monitor (GBM) \citep{Veres_2022} and has generated broad community interest. 

It is also one of a very small number of GRBs with detected very high energy emission, with reports of photon energy reaching 18 TeV by the LHAASO group \citep{Huang_2022} and 251 TeV by Carpet-2 \citep{Dzhappuev_2022}. The detection of very energetic photons in the TeV range such as in the case of GRB~190114C \citep{Magic_2019} and now GRB~221009A, challenges our understanding of GRB physics. There is no firm consensus on the production mechanisms of these very high-energy photons in the range of GeV to TeV \citep[see further discussions in][]{Balaji_2023,Mirabal_2023}. \cite{Atteia_2022} have predicted that the next event like GRB~221009A  has a $10\%$ probability of being observed in the next 50 years.

There is an ongoing large follow-up campaign using various ground-based telescopes covering the entire electromagnetic spectrum for GRB~221009A. Here, we present an extensive search for SN emission in GRB~221009A, focusing on the optical bands. First, we present the observations in Section~\ref{sec:obs}. Results from photometric and spectroscopic observations are provided in Section~\ref{sec:results}. Finally, we discuss the implications of these observations, along with concluding remarks, in Section~\ref{sec:con}. Throughout this paper, we use UTC date and time. We assume a $\Lambda$ Cold Dark Matter Universe with $H_0 = 70~\mathrm{kms}^{-1} \mathrm{Mpc}^{-1}$, $\Omega_m = 0.286$, and $\Omega_\lambda = 0.714$ \citep{Bennett_2014}. Presented uncertainties are at the $1\sigma$ confidence level. 

The extinction towards GRB~221009A is high, as it is at a Galactic latitude of $b$=4.3$^{\circ}$ and may also include a host component and/or a component associated with the local environment of the explosion. The assumed extinction value can affect interpretations of any underlying supernova component of the GRB. We derive an $E(B-V) = 1.32 \pm 0.06$ mag, representing a Milky Way-only extinction scenario \citep[using][]{Schlafly_2011}. We note that studies have shown the extinction values from the maps of \citet{Schlegel_1998} which \citet{Schlafly_2011} is based on are unreliable for areas near the Galactic plane \citep{Popowski_2003}. Recently \citet{Kann_2023} found that the value of the extinction can be lower than the value predicted from \citet{Schlafly_2011} using the method from \citet{Rowles_2009}. From their analysis, \citet{Kann_2023} concluded that the extinction towards the GRB~221009A is $0.709 < E(B-V) < 1.32$ mag.

However, \citet{Fulton_2023} analyzed the optical data and calculated a spectral index. They found that even after correcting for the Galactic extinction from \citet{Schlafly_2011} the optical and X-ray data do not agree. In order to address the discrepancy, they calculated the additional extinction required. This additional extinction changes with time and their quoted average is $A_v = 5.3$. One of the reasons for additional extinction was attributed to host galactic extinction.

For our analysis in this paper we consider two extinction values of $E(B-V) = 1.32 \pm 0.06$ mag and  $E(B-V) = 1.74$ mag. Throughout the paper, we assume $R_v= 3.1$, which translates into $A_v = 4.1$ and $A_v = 5.4$, respectively. We chose the simple case of Galactic extinction from \citet{Schlafly_2011} which is an upper limit from \citet{Kann_2023} analysis. In order to account for additional host galactic extinction, we also use the higher value reported by \citet{Fulton_2023}. This range of extinction is important for the search for an associated supernova as the higher extinction can hide the emerging supernova features.

\section{Observations and Data Reduction} \label{sec:obs}
\begin{figure}
\includegraphics[width=\columnwidth]{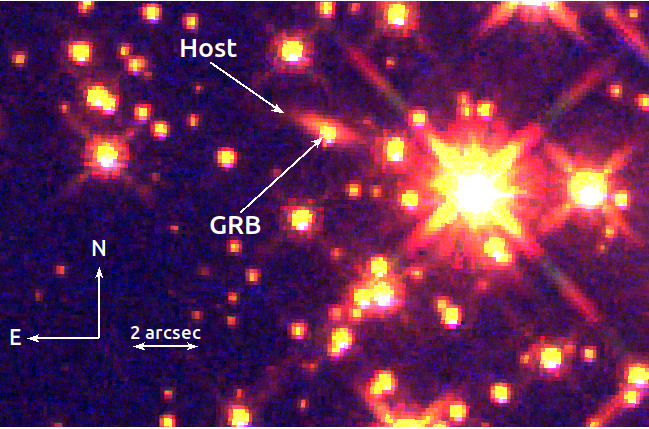}
\caption{Combined HST images WFC3/UVIS and WFC3/IR (F625W, F125W, F160W) of the GRB~221009A field, observed on 2022 Dec.~04.   Note the clear appearance of an underlying host galaxy, with a disk-like morphology.  GRB~221009A is slightly offset from the center of the apparent host, off the disk plane.
\label{fig:hst_image}}
\end{figure}
Following the discovery of GRB~221009A (Fig.~\ref{fig:hst_image}), we started an extensive ground-based follow-up campaign with the aim of detecting the SN associated with the GRB.  We augment this dataset with public archival data from the Hubble Space Telescope (HST) and the results reported to the GCN service.

\subsection{Photometry} \label{sec:obs_phot}

\begin{figure*}
\includegraphics[width=\textwidth]{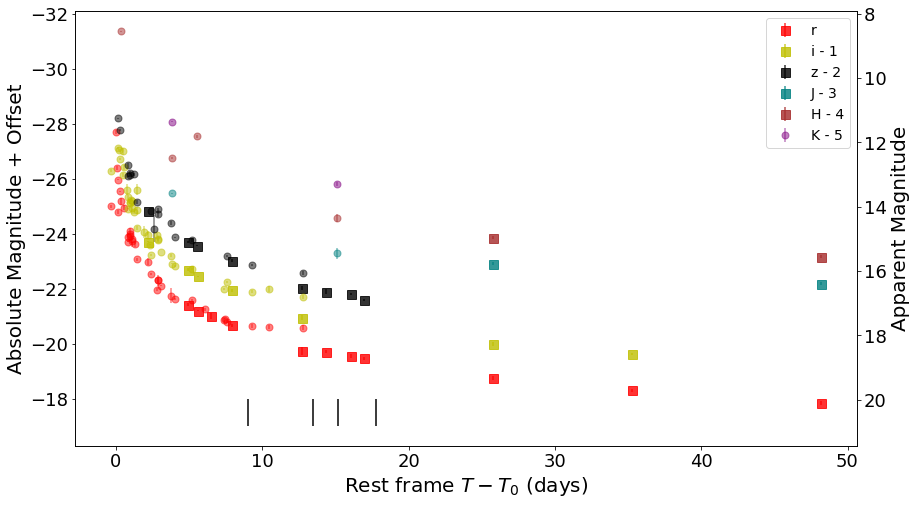}
\caption{ Optical light curve data in $r,i,z,J,H,K$ filters from GCN (in circles) and data reduced by our group (shown in squares). These values are from Tables~\ref{tab:photometry}, \ref{tab:hst_photometry} and \ref{tab:gcn} and corrected for Galactic extinction. The vertical black lines correspond to the time of our spectral observations.
\label{fig:mag_photometry}}
\end{figure*}

In Fig.~\ref{fig:mag_photometry} we present the photometric light curve of GRB~221009A from our own observations along with optical/NIR data from GCN circulars and publicly available data from HST.

\subsubsection{Ground-based Optical Observations}
We performed optical photometric follow-up using the following instruments: the Multi-Object Double CCD Spectrographs/Imagers \citep[MODS;][]{MODS} on the Large Binocular Telescope (LBT; 8.4~m twin telescope at Mt. Graham, Arizona), MuSCAT3 \citep{muscat} on the Faulkes Telescope North \citep{FTN} (FTN; 2~m telescope at Haleakal\=a Observatory, Hawaii) using Global Supernova Project time, Binospec \citep{Binospec} on the MMT (6.5~m telescope at Mt. Hopkins, Arizona), and the Goodman High-Throughput Spectrograph \citep[GHTS;][]{soar} on the Southern Astrophysical Research Telescope (SOAR; 4.1~m telescope at Cerro Pachón, Chile). Finally, GRB 221009A was observed by the Dark Energy Camera \citep[DECam;][]{Flaugher_2015} instrument on the 4-m Blanco Telescope at CTIO as a target of opportunity as part of the DECam Local Volume Exploration (DELVE) survey  \citep{Delve1,Delve2}. The details of the instruments and exposure times used, along with the observed magnitudes in different filters, are presented in Table~\ref{tab:photometry}. 

Images from Las Cumbres Observatory and DECam were preprocessed using BANZAI \citep{Banzai} and the DECam Community Pipeline \citep{decam_pipeline}, respectively. We bias subtracted and flat-fielded the remaining images using Python. We then performed aperture photometry using the Astropy Photutils package \citep{Bradley_2019} and calibrated to the Pan-STARRS1 catalog \citep{Chambers_2016}.  Table~\ref{tab:photometry} and Fig.~\ref{fig:mag_photometry} present all of our optical and NIR photometry.

\subsubsection{HST data}
We obtained publicly available data from Mikulski Archive for Space Telescopes (MAST) on GRB~221009A taken with the Wide-Field Camera 3 \citep[WFC3; ][]{WFC3_2022} UVIS and IR channel on HST on three epochs, 2022-11-08, 2022-11-19, and 2022-12-04, and measured photometry on the calibrated and combined optical and NIR images (Program 17264, PI: A. J. Levan). In Table~\ref{tab:hst_photometry}, we present our HST photometric results and other information. Photometry for the first epoch of HST data were also circulated via gamma-ray burst coordinates network (GCN) by \citet{Levan_2022}; our measurements are in good agreement within the respective errors. We assume the HST filters F625W, F775W, F098M, F125W, and F160W correspond to $r,i,y,J,H$ filters respectively. The results from our reductions are included in Fig.~\ref{fig:mag_photometry}. We note that the last two HST NIR data points could be contaminated due to the host galaxy.

\begin{deluxetable*}{cccccCCCCc}[t]
\tablewidth{0pt}
\tablecaption{GRB~221009A ground-based photometric data}\label{tab:photometry}
\tablehead{\colhead{Date (UT) } & \colhead{MJD} & \colhead{Facility} &  \colhead{Instrument} & \colhead{$t-t_0$ (\rm days)} &\colhead{Exposure (s)}&\colhead{r (mag)}&\colhead{i (mag)}&\colhead{z (mag)} & \colhead{GCN}}
\startdata
2022-10-12 & 59864.11 & LBT &   MODS  & 2.5 &6 \times 120 & \nodata & 19.00 \pm 0.02& 18.26 \pm 0.01 & 32759$^{a}$ \\
2022-10-15&59867.27 & FTN &  MuSCAT3 & 5.7 &3 \times 300& 21.13 \pm 0.06 & 20.05 \pm 0.05 & 19.39 \pm 0.05 & 32771$^{b}$\\
2022-10-15 & 59867.99 & CTIO & DECam & 6.4 & 1 \times 180 & 21.34 \pm 0.03 & 20.25 \pm 0.02 & 19.52 \pm 0.02 & \nodata \\
2022-10-17&59869.03 & SOAR &  GHTS & 7.4 & 3 \times 180&  21.53 \pm 0.10  &  20.42 \pm 0.10  &  19.75\pm 0.12 & \nodata \\
2022-10-18&59870.75 & FTN &  MuSCAT3 & 9.2 &3 \times 300& 21.88 \pm 0.06 & 20.77 \pm 0.04 & 20.06 \pm 0.04 & \nodata \\
2022-10-24&59876.20 & FTN &  MuSCAT3 & 14.6 &3 \times 300& 22.82 \pm 0.15 & 21.76 \pm 0.10 & 21.07 \pm 0.07 & \nodata \\
2022-10-26 & 59878.10 & MMT & Binospec  & 16.5 & 16 \times 30& 22.83 \pm 0.07 & \nodata & 21.19 \pm 0.08 & \nodata \\
2022-10-28 & 59880.07 & MMT & Binospec  & 18.5 & 16 \times 30& 23.00 \pm 0.07 & \nodata & 21.29 \pm 0.04 & \nodata \\
2022-10-29 & 59881.07 & MMT & Binospec  & 19.5 & 16 \times 30& 23.08 \pm 0.06 & \nodata & 21.50 \pm 0.04 & \nodata \\
\enddata
\tablenotetext{}{No correction for Galactic extinction has been applied}
\tablenotetext{a}{\citep{Shrestha_2022a}}
\tablenotetext{b}{\citep{Shrestha_2022b}}
\end{deluxetable*}

\subsubsection{GCN}
We collected all optical and NIR data reported via the GCNs for GRB~221009A, and optical data are presented in Table~\ref{tab:gcn}. As we are looking for subtle changes in brightness, consistent filter throughputs and photometric systems are essential to this analysis. For this reason, we decided to only make use of data that have been reported in AB magnitudes and calibrated to the Pan-STARRS1 catalog \citep{Chambers_2016}, for consistency. 
We corrected these data for Galactic extinction ($E(B-V) = 1.32$ mag and $E(B-V) = 1.74$ mag) before performing the data analysis.  

\subsection{Spectroscopy}\label{sec:obs_spec}

We took our first optical spectrum of GRB~221009A 10.4 days after the Burst Alert Telescope (BAT) trigger with the 10~m Hobby-Eberly Telescope \citep[HET;][]{het1, het2, het3} using the red arm of the Low-Resolution Spectrograph-2 (LRS2-R) having a $12'' \times 6''$ integral field unit and covering the 6450--8470 \AA\ wavelength interval with a resolution of $R \sim 1800$ \citep{Chonis16}. Sky subtraction and  wavelength and flux calibrations were performed by applying the 
Panacea\footnote{\url{https://github.com/grzeimann/Panacea}} pipeline implemented at HET. Further details on the instrument configuration and the standard data reduction steps can be found in \citet{Yang20}.

Furthermore, we observed the location of GRB~221009A using the Binospec imaging spectrograph on the MMT \citep{Fabrican_2019} for three different epochs. We used a long slit with a width of $1''$. For the observations done on 2022-10-25 (+15.48 days) and 2022-10-30 (+20.47 days), the 270 line mm$^{-1}$ grating was used with central wavelengths of 6500 \AA{} and 7500 \AA{}, respectively, as presented in Table~\ref{tab:spec}. For the 2022-10-27 (+17.49 days) observation, we used the 600 line mm$^{-1}$ grating with a central wavelength of 7500 \AA{}. We performed the initial data processing of flat-fielding, sky subtraction, and wavelength and flux calibration using the Binospec IDL pipeline \citep{BinoIDL}.\footnote{\url{https://bitbucket.org/chil_sai/binospec/wiki/Home}} 
1D spectra were extracted employing standard procedures using the Image Reduction and Analysis Facility (IRAF \footnote{IRAF \citep{iraf1,iraf2} was distributed by the National Optical Astronomy Observatory, which was managed by the Association of Universities for Research in Astronomy (AURA) under a cooperative agreement with the National Science Foundation.}). To account for slit losses, we scaled the spectrum to match the photometric data of GRB~221009A using a Python routine from \cite{hosseinzadeh_2022}. Finally, we corrected the spectrum for Galactic extinction using the Python \texttt{extinction} package \citep{extinction}.

A log of our spectroscopic observations can be seen in Table~\ref{tab:spec}, and the spectra can be seen in Figure~\ref{fig:spectroscopy} after correction for extinction and subtraction of an afterglow component as discussed in Section~\ref{subsec:spec}

\begin{deluxetable*}{ccccCCCCC}[t!]
\centering
\tablecaption{HST public data of GRB~221009A.  }\label{tab:hst_photometry}
\tablehead{\colhead{Date (UT) } & \colhead{MJD} &  \colhead{$t-t_0$ (\rm days)} &\colhead{Instrument} &\colhead{F625W}&\colhead{F775W}&\colhead{F098M} & \colhead{F125W} & \colhead{F160W}}
\startdata
2022-11-08 & 59891.27 & 29.68 & WFC3 & 23.79 \pm 0.32 & 22.72 \pm0.26 & 21.10 \pm 0.08 & 20.49 \pm 0.05 & 20.14 \pm 0.05 \\
2022-11-19 & 59902.18 & 40.59 & WFC3 &  24.24 \pm 0.51 & 23.08 \pm 0.40 & \nodata & \nodata & \nodata\\
2022-12-04 & 59917.06 & 55.46 & WFC3 &24.70 \pm 0.38 & \nodata & 21.82 \pm 0.11 & 21.21 \pm 0.07& 20.83 \pm 0.07\\
\enddata
\tablenotetext{}{No correction for Galactic extinction has been applied. The F625W and F775W filters are from WFC3/UVIS, while the F098M, F125W, and F160W filters are from WFC3/IR.}
\end{deluxetable*}

\vfill
\begin{deluxetable*}{cccCCCCCCc}[t!]
\tabletypesize{\scriptsize}
\tablecaption{GCN Photometric data of GRB~221009A.}\label{tab:gcn}
\tablehead{\colhead{GCN} & \colhead{Telescope} & \colhead{$t-t_0$(days) } & \colhead{r (mag)} & \colhead{i (mag)} & \colhead{z (mag)} & \colhead{J (mag)} &\colhead{H (mag)} &\colhead{K (mag)} & \colhead{Ref}}
\startdata
32647 & NEXT & 0.0087 & 14.93 \pm 0.05 & \nodata & \nodata & \nodata   & \nodata & \nodata  &\cite{Xu_2022}\\
32645  & AZT-33IK          & 0.01223     & 14.84 \pm 0.09 & \nodata     & \nodata   & \nodata & \nodata &\nodata & \cite{Belkin_2022}                         \\
32662  & GIT            & 0.094       & 16.16 \pm 0.07 & \nodata         & \nodata    & \nodata & \nodata &\nodata& \cite{Kumar_2022}                           \\
32646  & MeerLICHT      & 0.17        & 17.76 \pm 0.08 & 15.58 \pm 0.03 & 14.89 \pm 0.03&  \nodata & \nodata &\nodata & \cite{deWet_2022}\\
32644  & BOOTES-2/TELMA & 0.19        & 16.57 \pm 0.02 & \nodata     & \nodata   & \nodata   & \nodata &\nodata & \cite{Hu_2022}                       \\
32638  & LT             & 0.29        & 17.00 \pm 0.03    & 15.98 \pm 0.03 & 15.32 \pm 0.03    & \nodata & \nodata &\nodata& \cite{Perley_2022}\\
32755 & REM & 0.4 & \nodata & \nodata & \nodata & \nodata & 12.62 \pm 0.02 & \nodata & \cite{Davanzo_2022} \\
32652  & REM            & 0.434      & 17.36 \pm 0.12 & \nodata      & \nodata   & \nodata    & \nodata &\nodata&     \cite{Brivio_2022}                  \\
32664  & CDK  & 0.54      & \nodata                            & 15.70 \pm 0.13  & \nodata    & \nodata   & \nodata & \nodata &    \cite{Romanov_2022b}    \\
32659  & LOAO           & 0.63        & 17.55 \pm 0.06 & 16.41 \pm 0.05 & \nodata    & \nodata   & \nodata & \nodata &   \cite{Paek_2022}                   \\
32669  & Nickel           & 0.65 & 17.6 \pm 0.1   & 16.3 \pm 0.1   &   \nodata   & \nodata   & \nodata & \nodata &        \cite{Vidal_2022}             \\
32730  & Mitsume        & 0.9320      &    \nodata        & 17.1 \pm 0.2                  &  \nodata   & \nodata   & \nodata & \nodata &    \cite{Sasada_2022}      \\
32729  & AZT-33IK          &  1.02      & 18.84 \pm 0.02                   & 17.8 \pm 0.02      & 16.99 \pm 0.02  & \nodata   & \nodata & \nodata &      \cite{Zaznobin_2022}         \\
32667  & SLT-40cm          & 1.04      & 18.67 \pm 0.16      & 17.38 \pm 0.09  & 16.60 \pm 0.09 & \nodata   & \nodata & \nodata &    \cite{Chen_2022}              \\
32795  & GRANDMA        & 1.1360      & 18.57 \pm 0.05     & 17.56 \pm 0.05    & 16.93 \pm 0.05    & \nodata   & \nodata & \nodata &  \cite{Rajabov_2022}           \\
32684  & Sintez-Newton  &   1.163     & 18.43 \pm 0.10      &            \nodata & \nodata      & \nodata   & \nodata & \nodata   &          \cite{Belkin_2022c}         \\
32670  & AZT-20           & 1.164         & 18.64 \pm 0.03 & 17.58 \pm 0.01    & 16.87 \pm 0.05   & \nodata   & \nodata & \nodata &    \cite{Kim_2022}            \\
32693  & Las Cumbres 1~m          &   1.22        & 18.80 \pm 0.21   & 17.8 \pm 0.2  &  \nodata    & \nodata   & \nodata & \nodata &     \cite{Strausbaugh_2022}   \\
32709  &RC80       & 1.26       & 18.74 \pm 0.12 & 17.5 \pm 0.08  &      \nodata      & \nodata   & \nodata & \nodata &    \cite{Vinko_2022}   \\
32678  & BG3-Opal      & 1.48       &      \nodata                        & 17.92 \pm 0.06                   & 16.92 \pm 0.05 &         \nodata   & \nodata & \nodata & \cite{Groot_2022}         \\
32679  & T24 iTelescope     & 1.66       & \nodata    & 17.1 \pm 0.2       &   \nodata            & \nodata   & \nodata & \nodata &  \cite{Romanov_2022}           \\
32705  & COATLI         &  2.63       &      \nodata                         & 19.10 \pm 0.02                    &   \nodata & \nodata   & \nodata & \nodata & \cite{Butler_2022}                         \\
33038 & FTN & 2.73 & 20.01 \pm 0.03 & 19.48 \pm 0.03 & 18.26 \pm 0.01 & \nodata   & \nodata & \nodata & \cite{Kimura_2022} \\
32727  & GMG            &  3.0        &    \nodata                          &        \nodata                       & 18.9   & \nodata   & \nodata & \nodata & \cite{Mao_2022}                   \\
32753 & T120cm & 3.25 & 20.23 \pm 0.09  & 18.91 \pm 0.11 & 18.35 \pm 0.13 & \nodata   & \nodata & \nodata & \cite{Schneider_2022}\\
32743  & RTT-150        & 3.3          & 20.24 \pm0.19  & 18.92 \pm 0.04& 18.10 \pm 0.04  & \nodata   & \nodata & \nodata & \cite{Bikmaev_2022}      \\
32738  & Las Cumbres 1~m         &    3.53      & $>$21              & $>$20.5            &    \nodata      & \nodata   & \nodata & \nodata &   \cite{Strausbaugh_2022b}                 \\
32739  & LDT         &  3.57       & 20.44 \pm 0.02                   & 19.37 \pm 0.01                   &                        & \nodata   & \nodata & \nodata & \cite{Connor_2022b}     \\
32752  & RTT-150        &  4.3      & 20.86 \pm 0.27 & 19.50 \pm 0.07  & 18.70 \pm 0.06 & \nodata   & \nodata & \nodata & \cite{Bikmaev_2022b} \\
32750 & Gemini & 4.4 & \nodata & \nodata & \nodata & 17.93 \pm 0.03 & 17.23 \pm 0.05 & 16.69 \pm 0.02 & \cite{oconnor_2022} \\
32749  & Gemini         &   4.437   &              \nodata                 & 19.8                          &  \nodata                         & \nodata   & \nodata & \nodata &   \cite{Rastinejad_2022} \\
32758 & Pan-STARRS1 & 4.67 & 20.92 \pm 0.05 & 19.88 \pm 0.02 & 19.21 \pm 0.02 & \nodata   & \nodata & \nodata & \cite{Huber_2022} \\
32769 & AZT-20 & 6.05 & 20.96 \pm 0.05 & 20.00 \pm 0.04 & 19.31 \pm 0.08 & \nodata   & \nodata & \nodata & \cite{Belkin_2022b} \\
32804 & TNG & 7.3&\nodata   & \nodata & \nodata   & \nodata &   16.45 \pm 0.04 &\nodata   & \cite{Ferro_2022} \\ 
32809  & LBT            &  8.58     & 21.63 \pm 0.02   &              \nodata &   \nodata   & \nodata & \nodata &\nodata  &        \cite{Rossi_2022}         \\
32799  & LDT            &   9.5      & 21.68 \pm 0.07  & 20.72 \pm 0.05  & \nodata & \nodata   & \nodata & \nodata & \cite{Connor_2022}\\  
32818 & AZT-20 & 12.05 & 21.94 \pm 0.07 & 20.72 \pm 0.11 & \nodata& \nodata   & \nodata & \nodata & \cite{Belkin_2022} \\
32860 & Gemini & 17.4 & \nodata & \nodata &\nodata & 20.1 \pm 0.2 & 19.43 \pm 0.15 & 18.94 \pm 0.08 & \cite{oconnor_2022b} \\
32921 & HST & 29.68 & 23.61 \pm 0.04 & 22.43 \pm 0.04 & \nodata& \nodata   & \nodata & \nodata & \cite{Levan_2022} \\
\enddata
\tablenotetext{}{No correction for Galactic extinction has been applied. }
\end{deluxetable*}

\begin{deluxetable*}{cccccccC}[t]

\tablecaption{Spectroscopic Observations}\label{tab:spec}
\tablehead{\colhead{UT Date} & \colhead{MJD} & \colhead{Facility} &  \colhead{Instrument} & \colhead{Filter}& \colhead{Grating} &\colhead{$t-t_0$ (days)} &\colhead{Exposure (s)}}
\startdata
2022-10-21 & 59873.10 & HET & LRS2-R & -- & VPH-Grism & 10.40 & 1 \times 2000 \\
2022-10-25&59877.07 & MMT &  Binospec & LP3800& 270 & 15.48 &4 \times 1200 \\
2022-10-27&59879.08 & MMT &  Binospec & LP3800& 600 & 17.49 &5 \times 1200 \\
2022-10-30&59882.06 & MMT &  Binospec & LP3800& 270 & 20.47 &4 \times 1200 \\
\enddata
\vspace{-3cm}
\end{deluxetable*}

\section{Supernova Limits} \label{sec:results}

The observation of SN 1998bw coincident spatially and temporally with GRB 980425 \citep{Galama_1998, Iwamoto_1998, Kulkarni_1998} was the first direct evidence of the GRB-SN association. For nearby (z$<$0.3) long GRBs, such as GRB~221009A, there have been observations of an associated SN \citep{Hjorth_2012} in all except four cases: GRB~060505 \citep{Fynbo_2006}, GRB~060614 \citep{Fynbo_2006,Gal_2006,Della_2006}, GRB~111005A \citep{Michal_2018,Tanga_2018}, and GRB~211211A \citep{Rastinejad_2022_kilonova, Troja_2022}. However, the true fraction of long GRBs without a SN is unknown \citep{Hjorth_2012}. The study of associated SNe provides an important clue to the understanding of the progenitors and environment of these long GRBs. Thus, we search for supernova signatures in our GRB~221009A light curves and spectra in Subsection~\ref{subsec:lc} and Subsection~\ref{subsec:spec}, respectively.

\subsection{Light Curve Analysis} \label{subsec:lc}

For our light curve analysis, we focused on the $r$ and $i$ bands because we have better data coverage in these bands compared to the $z$ band. The emission observed at the GRB position will have contributions from three potential components: the GRB afterglow, the supernova, and the host galaxy. The host galaxy is detected in the WFC3/IR filters whereas, in the WFC3/UVIS filters, it is not clearly visible as shown in  Figure~\ref{fig:hst_image}. Therefore we do not explicitly model its contribution for our optical analysis. In addition, \citet{Levan_2023} recently presented a detailed analysis of the GRB host galaxy and they found the host extinction to be $A_V = 0.019^{+0.030}_{-0.014}$, which points to low host galaxy extinction, at least in some analyses.
That said, we do consider two extinction scenarios as discussed earlier, with one focused solely on Milky Way extinction ($E(B-V)$=1.32) and another which may include a host galaxy contribution ($E(B-V)$=1.74; \citealt{Fulton_2023}). We approached the search for the associated SN component using two different techniques.

\begin{figure*}
\includegraphics[width=\textwidth]{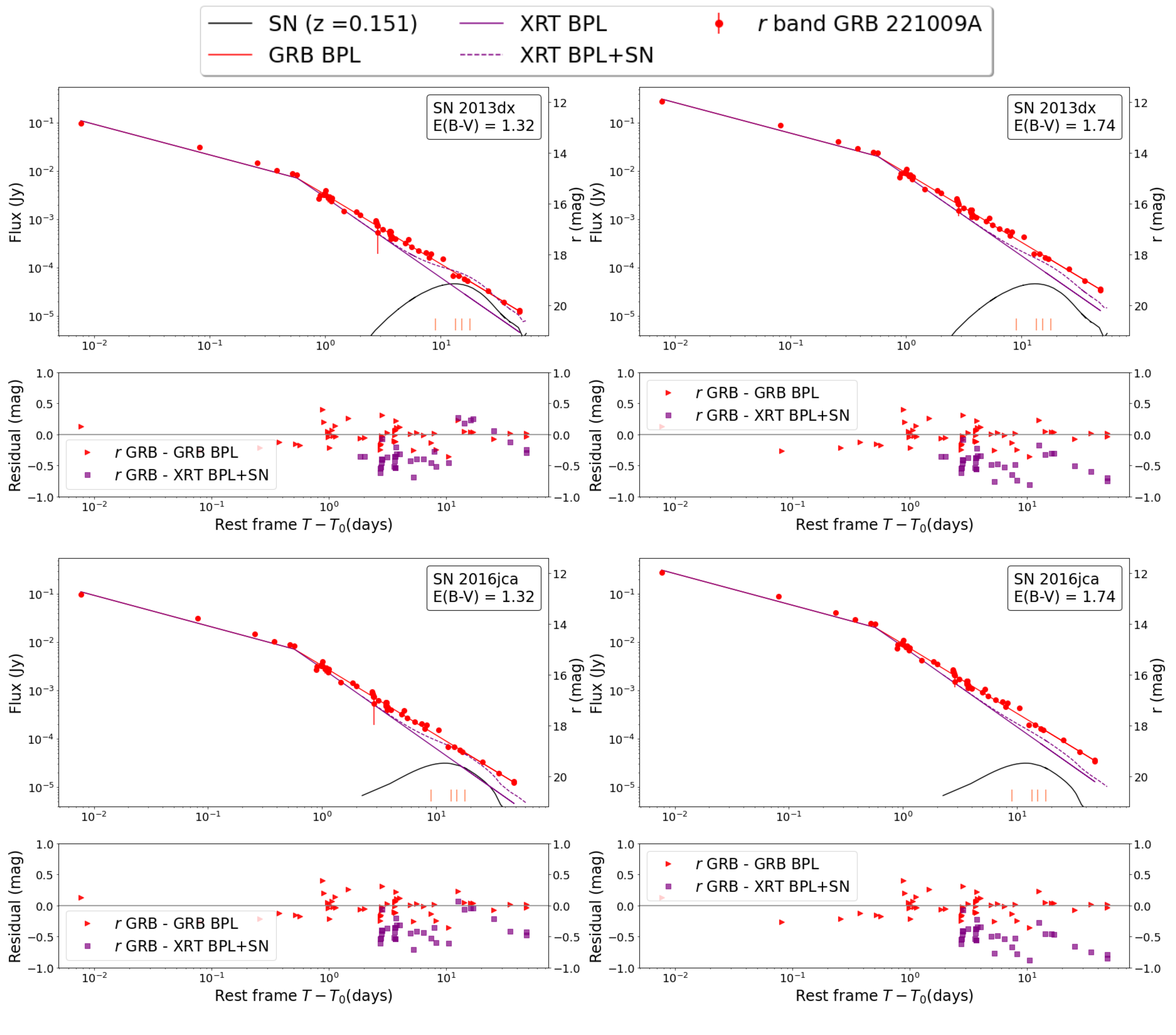}
\caption{Light curves for the $r$ filter for two different extinction values of $E(B-V) =1.32$ mag (left) and $E(B-V) =1.74$ mag (right)  with a broken power-law fit to the data. Light curves models of SN~2013dx (top) and SN~2016jca (bottom) are included as black solid lines along with GRB data. For each case, a residual plot in magnitude is presented. The purple square points are the difference between supernova contribution plus GRB afterglow estimated by the XRT decay index and broken-power-law estimate. The red triangle points are the difference between the observed GRB magnitude and broken-power-law estimate. The orange lines in the figure are the epochs we have spectroscopic observations. The residual plot shows that the SN component is not necessary to explain the light curve.
\label{fig:lc_models_r_XRTSN}}
\end{figure*}

\begin{figure*}
\includegraphics[width=\textwidth]{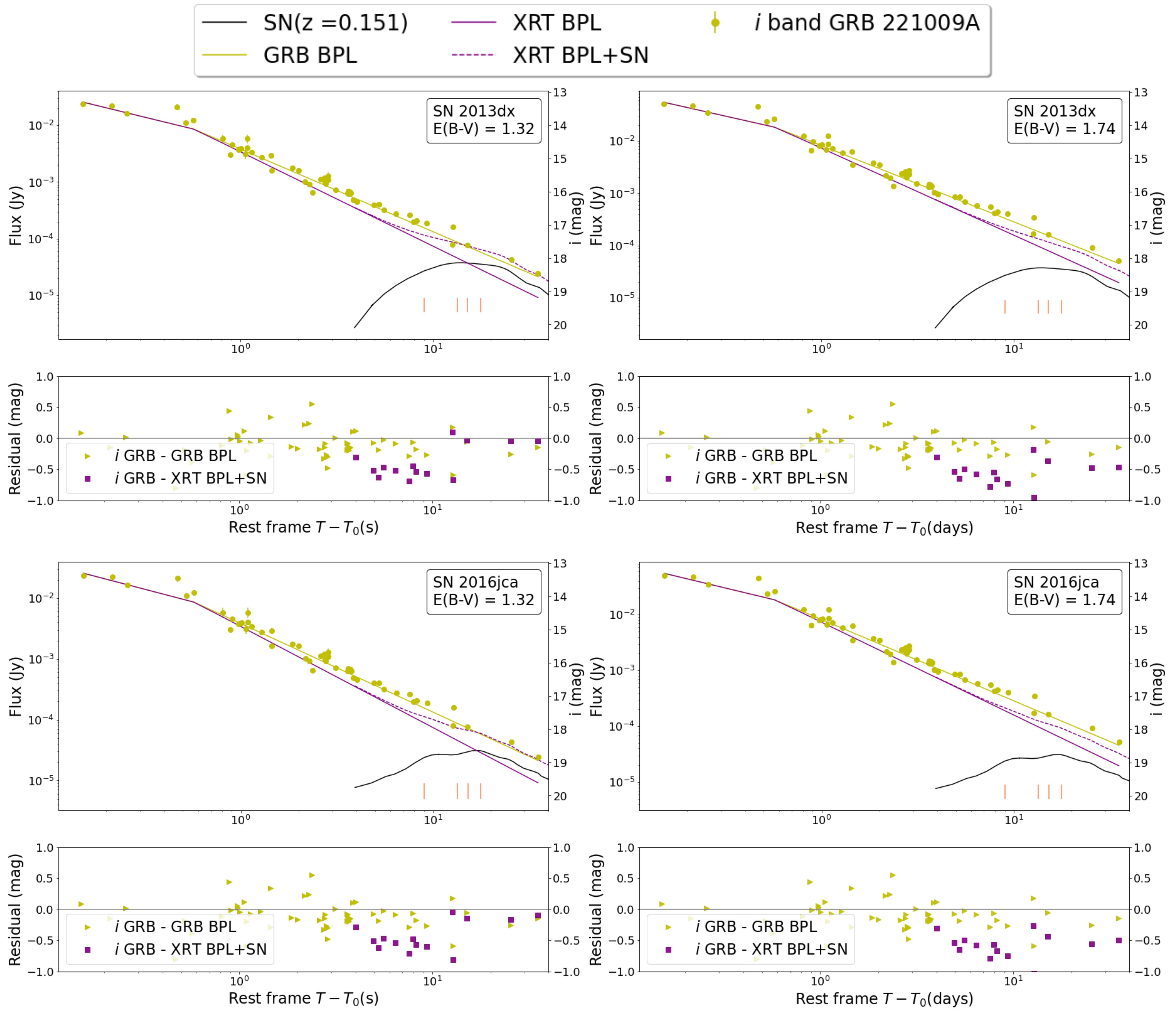}
\caption{ Same plot as Fig.~\ref{fig:lc_models_r_XRTSN} for $i$ filter. The yellow triangle points are the difference between the observed GRB magnitude and broken-power-law estimate. The residual plot shows that the broken power-law fit to the GRB optical data is the best fit for the observed data.
\label{fig:lc_models_i_XRTSN}}
\end{figure*}

For the first technique, we model the GRB optical afterglow using a broken power law model, assuming the decay index from the XRT data to avoid possible supernova contamination in the optical data \citep[e.g.][]{Toy_2016, Fulton_2023}. The reduced $\chi^2$ for a one-break fit of the XRT data is 1.33 whereas for four breaks (the best-fit model) is 1.24. We also calculated the Bayesian information criterion: $BIC = kln(n) -  2 ln (L)$  \citep{Kass_1995}, where $k$ is the number of fit parameter, $n$ is the number of data points, $L$ is the estimation of the likelihood at its maximum. We find the ratio of BIC for the 4 break model to 1 break model to be 9.9 which shows that there is strong evidence against the higher break model \citep{Kass_1995}.
Hence, we do not find a significant improvement in the fit beyond one break and thus use a one-break model for the remainder of the analysis. The one-break fit of the XRT data has a decay index of $\alpha_1 = 1.515 \pm 0.003$ before the break time of $0.6$ days and after the break the decay index is $\alpha_2 = 1.663 \pm 0.006$\footnote{\url{https://www.swift.ac.uk/xrt_live_cat/01126853/}}  \citep{Evans_2009}. The solid purple line represents this decay index in Fig.~\ref{fig:lc_models_r_XRTSN} and Fig.~\ref{fig:lc_models_i_XRTSN}. Hence, we force the decay index after $0.6$ days for optical data to be $1.663 \pm 0.006$ which is the $\alpha_2$ and are labeled `XRT BPL' in Fig.~\ref{fig:lc_models_r_XRTSN} and Fig.~\ref{fig:lc_models_i_XRTSN}.

We also fit all the optical data points with an empirical broken power law, whose best-fit is represented by the solid red and yellow lines for the $r$ and $i$ filters, respectively. The best-fit model to the optical data gives the decay index values of $0.64$ before and $1.44$ after the break time of $0.6$ days for the $r$ band data. For the $i$ band data, the best-fit model has decay index values of $0.81$ before and $1.46$ after the break time of $0.6$ days. These broken power-law fits to the optical data are labeled `GRB BPL' in Fig.~\ref{fig:lc_models_r_XRTSN} and Fig.~\ref{fig:lc_models_i_XRTSN}. In addition, we also fit a broken power law to limited data points in $z$ band which are not shown here. The best fit gives decay indices of $0.97$ and $1.38$ before and after the break time of $0.6$ days respectively which is consistent with $r$ and $i$ band data.

In addition to the power-law fits above, we consider an additional supernova component using data directly from two supernovae associated with GRBs: SN~2013dx/GRB~130702A \citep{Toy_2016} and SN~2016jca/GRB~161219b \citep{Cano_2017b}. SN~2013dx ($M_{r,max}$=$-$19.54) is at a redshift of 0.145 and SN~2016jca ($M_{r,max}$=$-$19.04) is at a redshift of 0.1475, both of which are similar to the redshift of GRB~221009A; these objects were specially chosen so that we could do direct filter comparisons between them and GRB~221009A without any $K$ correction. These comparisons in the $r$ and $i$ bands are presented in Fig.~\ref{fig:lc_models_r_XRTSN} and Fig.~\ref{fig:lc_models_i_XRTSN} respectively. Each row in this figure represents a model for one supernova, where the black solid line is the supernova light curve. We note that \citet{Mazzali_2021} and \citet{Ashall_2019} have also carried out analysis of SN~2013dx and SN~2016jca respectively. From their analysis they found the luminosities to be different from the models we considered in our analysis.

In order to determine the effect of the supernova on the light curve of GRB~221009A qualitatively we combined the afterglow component (purple solid line) with the supernova component (black solid line). The resulting light curve is shown as a dashed line in the plot. For the case with a lower extinction value (left panels of Fig.~\ref{fig:lc_models_r_XRTSN} and Fig.~\ref{fig:lc_models_i_XRTSN}), the dashed lines initially underpredict the flux compared to the observed flux. This can be also seen in the residual plots shown in the bottom panels. During the SNe peak, the match to the observed data is better. However, in the $r$ band we find an excess in observed flux compared to the X-ray power law + SN model. For the case of $E(B-V) = 1.74$ mag, the dashed lines underpredict the flux compared to observed data at all times. The residual plots for all the cases show that the broken power law fitted to optical data produces the least residual, without any need for a SN component.



\begin{figure*}
\includegraphics[width=\textwidth]{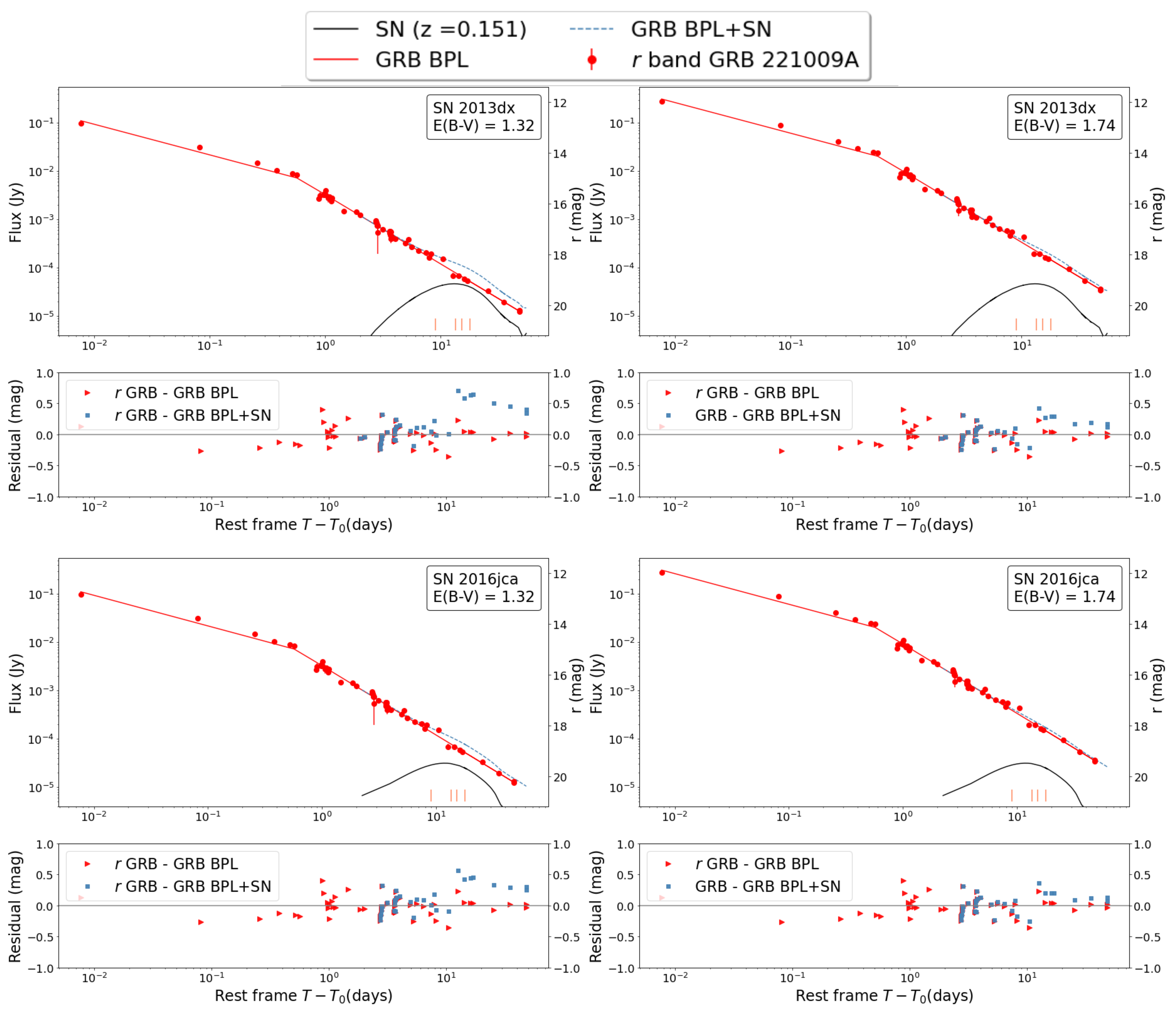}
\caption{ Light curves for the $r$ filter using the two different extinction values of $E(B-V) =1.32$ mag (left) and $E(B-V) =1.74$ mag (right). SN~2013dx and SN~2016jca light curves are plotted as black solid lines in top and bottom panels respectively. The sum of SN contribution and GRB afterglow is shown as dashed blue lines. For each case, a residual plot in magnitude is presented. The blue squares are the difference between the supernova contribution plus GRB afterglow estimated by a broken-power-law fit to the optical data. The red triangle points are the difference between the observed GRB magnitude and the broken-power-law estimate. 
\label{fig:lc_models_r_GRBSN}}
\end{figure*}

\begin{figure*}
\includegraphics[width=\textwidth]{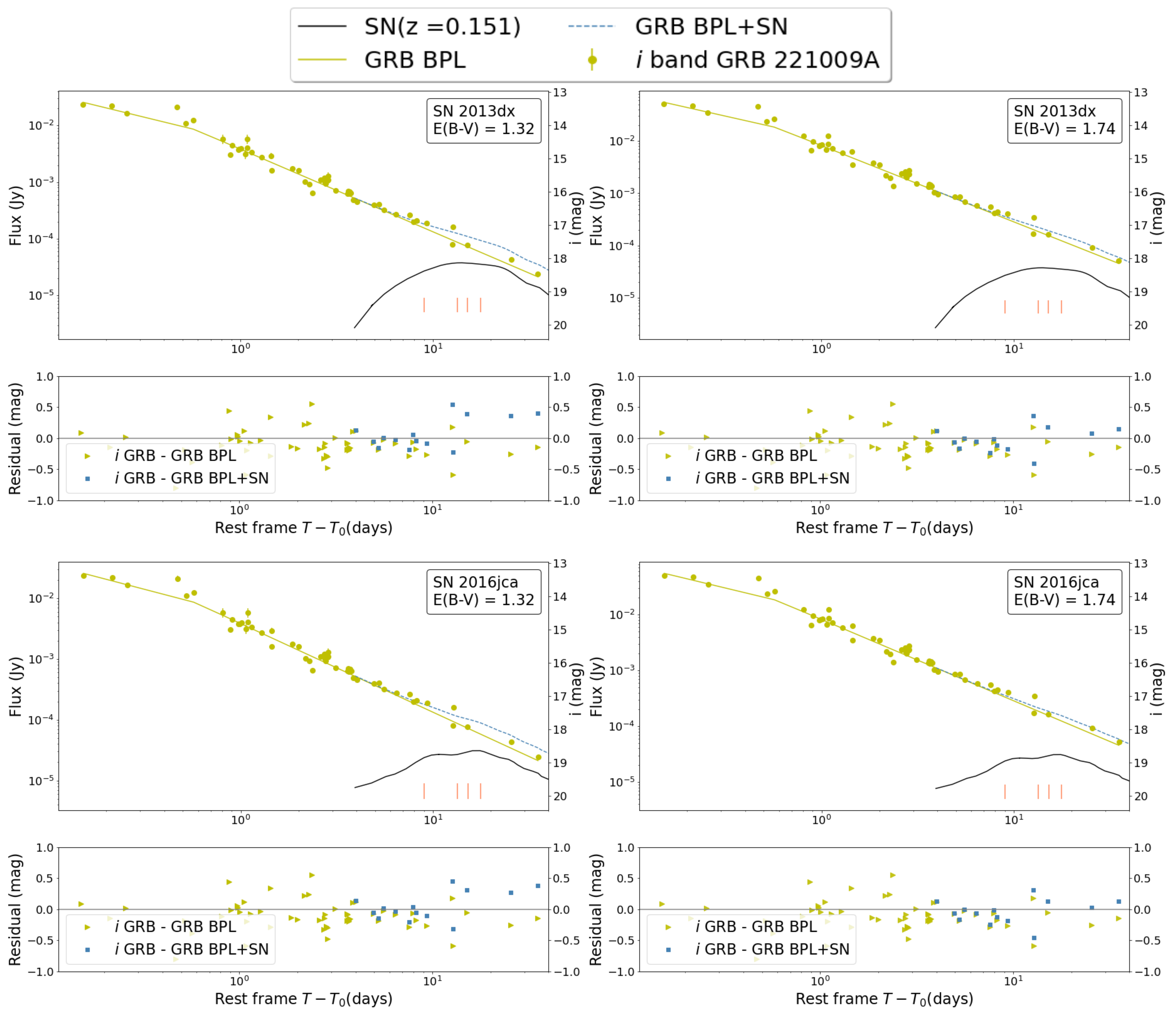}
\caption{ Same plot as Fig.~\ref{fig:lc_models_r_GRBSN} for the $i$ filter. The blue squares are the difference between supernova contribution plus GRB afterglow estimated by broken-power-law fitted to optical data. The yellow triangle points are the difference between the observed GRB magnitude and the broken-power-law estimate. 
\label{fig:lc_models_i_GRBSN}}
\end{figure*}

For the second method, we assume that the optical light curve is dominated by the GRB afterglow and it is defined by the broken power law fit to the optical data. As seen in the earlier discussion, this gives the least residual. In addition, the optical decay index in $r$ band after the break is $\alpha_{opt,r} = 1.44$ and for $i$ band is $\alpha_{opt,i} =1.46$. The decay index for XRT is $\alpha_{X} = 1.663$. We see that the difference between the optical and X-ray decay indices is close to $1/4$, which is expected for the slow cooling regime for constant interstellar medium for the case where the synchrotron cooling break is between the optical and the X-ray band \citep{Zaninoni_2013}. \citet{Laskar_2023} show that an afterglow model with a wind-like density profile can also match the optical and X-ray light curves without the addition of a SN component. In their model, the cooling break is above the X-ray band and the characteristic synchrotron frequency (corresponding to the minimum electron energy) is just below the optical band, which causes the optical light curves to decline more slowly than the X-rays. With this assumption, we investigated the effect of an associated SN in the light curve. We add the contribution of SN~2013dx and SN~2016jca to the GRB afterglow model with broken power law and decay index of $1.44$ and $1.46$ after the break for $r$ and $i$ bands respectively. For the case of $E(B-V) =1.32$ mag, we see that both in $r$ (Fig.~\ref{fig:lc_models_r_GRBSN}) and $i$ (Fig.~\ref{fig:lc_models_i_GRBSN}) bands, a supernova bump would have been clear for both the SNe cases. However, if the $E(B-V)$ is high as 1.74 mag, then the bump is not clear in the light curves and it could be hidden by the afterglow. This gives us a qualitative limit of $M_{r,max} \approx-$19.5 mag.

\subsection{Spectroscopic Analysis} \label{subsec:spec}
\begin{figure*}
\includegraphics[width=\textwidth]{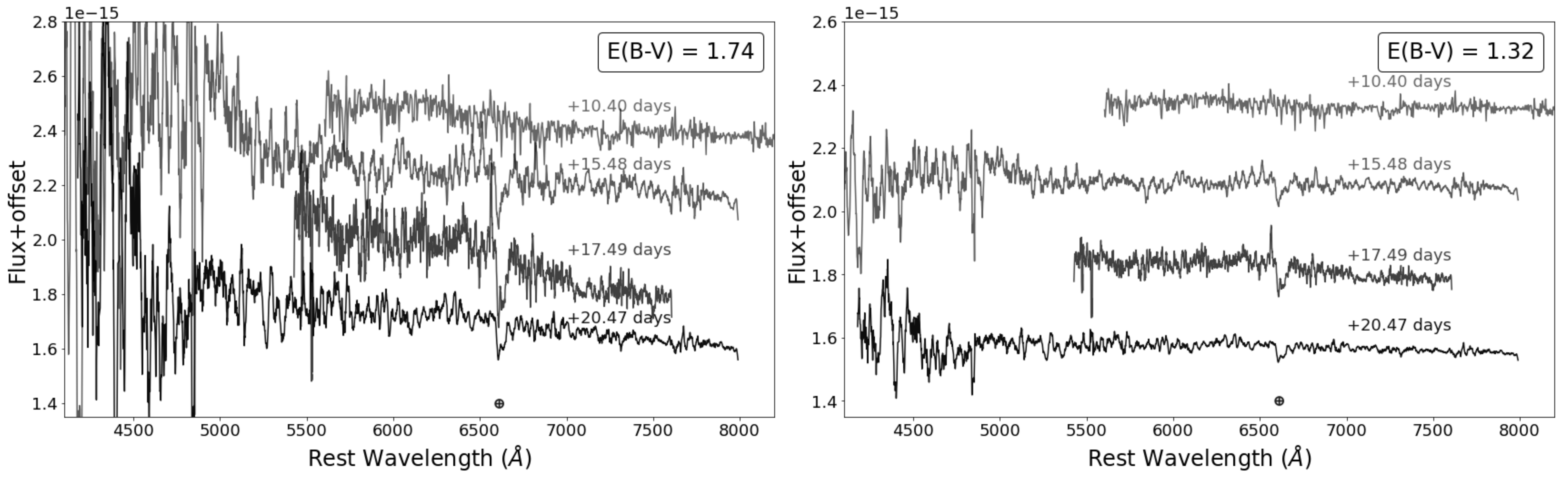}
\caption{Smoothed spectra of GRB~221009A observed at four different phases of +10.4 (HET), +15.48 (MMT), +17.49 (MMT), and +20.47 (MMT) days after the trigger using two different extinction values of $E(B-V)$=1.32 mag (left) and $E(B-V)$=1.74 mag (right). GRB afterglow contributions have been subtracted from the presented spectra. The wavelengths have been converted to the rest frame assuming a redshift of 0.151. No clear SN Ic-BL features are seen in the spectra.
\label{fig:spectroscopy}}
\end{figure*}

We observed the location of GRB~221009A spectroscopically using the HET and MMT. The observations were taken at four different epochs to detect the broad spectral features which are observed in type SNe Ic-BL associated with GRBs. In Fig.~\ref{fig:spectroscopy}, we present the results of spectroscopic observations done at the location of GRB~221009A on four different epochs using HET and MMT after smoothing. The spectra have been corrected for two different extinction values of $E(B-V)= 1.32$ mag (left panel) and $E(B-V)= 1.74$ mag (right panel). We note tentative, narrow H$\alpha$ emission at $z$=0.151 in the +17.8 day spectrum, which may also be visible in the +10.4 day HET spectrum as seen by \citet{Izzo_2022}; we take this as tentative confirmation of the host redshift.

We also attempted to isolate the supernova features in these spectra by modeling the contribution of the GRB afterglow in the spectra. To model the GRB afterglow contribution, we used the analytic function $F_{\nu} \propto t^{-\alpha} \nu^{-\beta}$ where $F_{\nu}$ is the flux, $\nu$ is the frequency, $\beta$ is the spectral index, $t$ is the time since trigger, and $\alpha$ is the photometric decay index. We used our data in the $r$, $i$, and $z$ filters from day 6.4 and $E(B-V) = 1.32$ to fit a power law and calculate the spectral decay index. We found $\beta_{opt} = 0.59 \pm 0.17$, which we then used to calculate the flux. We note that our calculated spectral index does not follow $\beta_{X} = \beta_{opt} + 0.5$, where $\beta_X = 0.9$ is the spectral index from XRT data. We calculated $F_{\nu}$  for the times of each observed spectrum and we performed photometric calibration to extinction-corrected ($E(B-V)= 1.32$ mag and $E(B-V)= 1.74$ mag) broadband photometry. Finally, we subtracted this contribution from the original spectra to extract supernova features. We do not detect any clear supernova features such as a broad absorption feature of \ion{Si}{2} $\lambda 6355$ even after this correction in our spectra.  

\begin{figure*}
\centering
\includegraphics[width=0.9\textwidth]{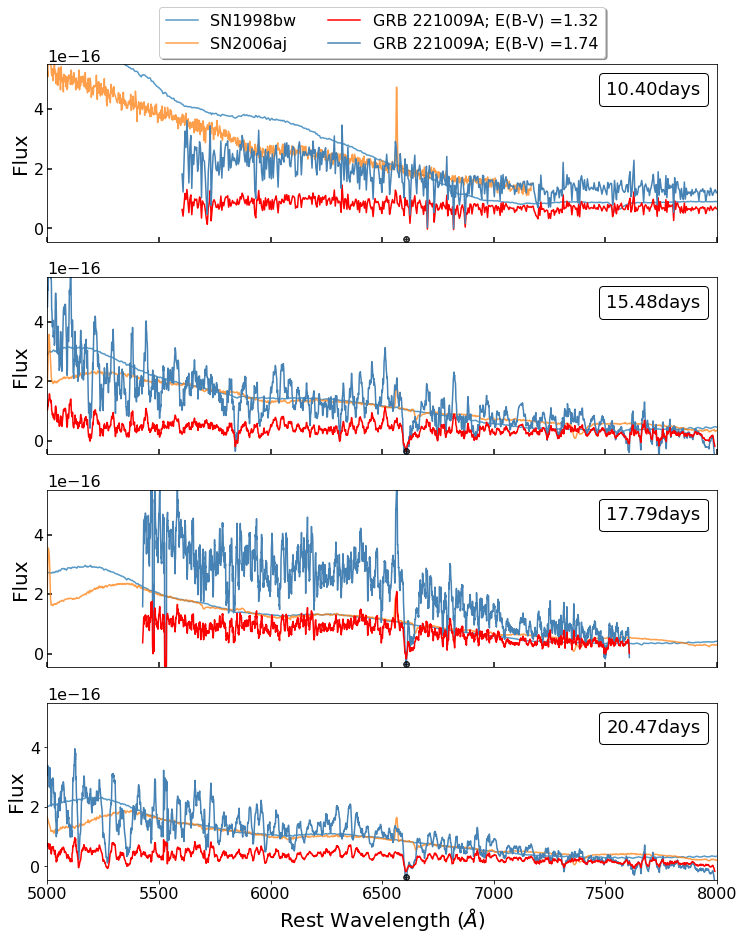}
\caption{ Smoothed spectra of GRB~221009A compared to two supernovae, SN~1998bw and SN~2006aj, which were both associated with GRBs. The spectra have been corrected for extinction using the two values used throughout this work ($E(B-V)$=1.32 mag and $E(B-V)$=1.74 mag) and have been corrected for their redshift appropriately. All the spectra are calibrated to the photometry of GRB~221009A at the same phase and the GRB afterglow contribution is subtracted. The four panels are for four different times since the GRB trigger time: 10.40 days, 15.48 days, 17.49 days,  and 20.47 days. The spectrum of GRB~221009A at 10.40 days is from HET, and the spectra at 15.48 days, 17.49 days, and 20.47 days are from the MMT. Although the initial HET spectrum bears some resemblance to SN~1998bw in the high extinction scenario, overall there are no clear supernova features present. 
\label{fig:spectroscopy_comp}}
\end{figure*}

We also compared the spectra of GRB~221009A to two different supernovae, SN~1998bw \citep{Patat_2001} and SN~2006aj, associated with GRB~980425 and GRB~060218 \citep{Modjaz_2006,Pian_2006} respectively. We selected these two cases because SN~1998bw is a bright SN associated with GRB where as SN~2006aj falls in the lower luminosity category. We obtained their spectra from the Weizmann Interactive Supernova Data Repository \citep{Yaron_2012} at similar phases to our spectra. The SN~1998bw spectra in WISeREP were from \citet{Patat_2001} and those for SN~2006aj were from \citet{Modjaz_2006, Pian_2006}. These spectra from the archive were first corrected for the Galactic extinction $E(B-V)$ of 0.0509 mag and 0.1253 mag for SN~1998bw and SN~2006aj, respectively. We also shifted the spectra from the redshifts of 0.0085 and 0.0331 for the two supernovae to the redshift of GRB~221009A of 0.151. Finally, these redshifted spectra were scaled to match to extinction-corrected photometry of GRB~221009A. The comparison between the calibrated spectra and the spectra of GRB~221009A at four different epochs is presented in Fig.~\ref{fig:spectroscopy_comp}. For both the SN~1998bw and SN~2006aj spectra, broad features indicative of SN Ic-BL type can be seen for all the epochs. However, the GRB~221009A spectra are noisier than the SN~1998bw and SN~2006aj spectra. At the first epoch i.e. 10.40 days after the GRB trigger, our HET spectrum when corrected for $E(B-V) = 1.74$ (blue line in Fig.~\ref{fig:spectroscopy_comp}) has a similar structure as the other two supernovae. However, there are no distinct broad absorption features as expected for SN Ic-BL. MMT spectra at 15.48 days and 17.79 days after the trigger do not show any clear features. For the last spectrum at 20.47 days after the trigger observed by MMT, there is a broad feature with some noise between 6000 and 6500 $\AA$ similar to the other two SNe. However, we do not see clear broad absorption features which are indicative of the SN component. We also note that \citet{Fulton_2023} reported that the peak of the supernova they find associated with the GRB~221009A is 20 days after the trigger, the same as the epoch of our last spectral observation.

\section{Discussions and Conclusion} \label{sec:con}
For this work, we observed the location of GRB~221009A both using imaging and spectroscopy to search for supernova signatures. We made our first observations using the LBT MODS imager starting 2.5 days after the BAT trigger and augmented our photometry with publicly available data from GCN and HST to do our photometric analysis. Our light curve modeling does not find a supernova bump in the light curve, as would be expected for a GRB with a supernova component, though we are unable to exclude the presence of a faint supernova for a high extinction case. In addition, we obtained spectroscopic observations from the HET and MMT. The spectra do not contain clear broad line features indicating associated SN Ic-BL. We summarize the results below:

\begin{enumerate}
    
    \item We investigate if the optical light curves contain contributions from the GRB afterglow and a possible supernova. We fit the light curves with two different models. The first model fixes the afterglow contribution to a broken power law with the same decay rate as the X-ray light curve (assumed to be dominated by the GRB afterglow at all times). The second model is a broken power law with a decay rate fit to match the observed data, which assumes that all of the emission is contributed by the GRB afterglow.  We also added a SN component at late times for both cases to check if the observed light curve could be the result of afterglow and SN emission for the first model and if a supernova component would have been identified for the second model.

    \item We checked how the light curve would evolve in the presence of two different supernovae at a similar redshift to GRB 221009A. In all cases, we find that the afterglow+supernova model provides a worse fit to the data than the afterglow-only model. Thus, we do not find evidence of a bright supernova in our light curve.
    \item We observed the spectrum of GRB~221009A to search for spectroscopic signatures of a supernova. We modeled the GRB afterglow contribution to the spectrum as $F_{\nu} \propto t^{-\alpha} \nu^{-\beta}$. We estimated the value of $\beta$ using optical photometry. Subtraction of this contribution does not change the features of the spectra. It only brings the flux values closer to zero. We do not detect any consistent, broad features across our spectroscopic sequence. Hence, no clear supernova contribution is detected.
\end{enumerate}

Our study shows that we do not detect a supernova associated with the GRB~221009A. However, we cannot discard the possibility that there could be an associated supernova below our detection limit if we consider the high extinction. The lack of supernova features could be due to the supernova being fainter than SN~2013dx or SN~2016jca (i.e., with an absolute magnitude of $M_r > -19.5$ and $M_i > -19.3$) or due to the high extinction in the direction of GRB~221009A, which could hide the supernova features.  Our non-detection of a supernova from GRB~221009A at this level may suggest that most of the energy produced by the central engine is carried by the relativistic jet, not the bulk ejecta. 

We also consider our results in the context of the small but growing sample of GRBs detected at very high gamma-ray energies (TeV). In the case of GRB~190114C, the first GRB with associated TeV photons, an associated supernova was also seen (SN~2019jrj; \citealt{Melandri_2022}). Interestingly, the spectral analysis of \citet{Melandri_2022} shows that SN~2019jrj's absorption lines are more similar to those of less luminous CCSNe such as SNe 2004aw \citep{Taubenberger_2006} or 2002ap \citep{Tomita_2006} than to other GRB-SNe. Specifically, SN~2019jrj's emission lines are narrower, meaning that its ejecta was more slowly moving and carried less energy. The photometric analysis of \citet{Melandri_2022} shows that the luminosity of SN~2019jrj is lower than SN~2013dx. This luminosity range would be below our detection limit if we assume the extinction to be 1.74 mag for GRB~221009A. In this case, our results suggest the possibility that GRBs capable of accelerating photons to TeV energies are accompanied by underluminous supernovae, possibly providing clues to their central engines and progenitors.

\citet{Fulton_2023} reported the detection of a supernova component in their optical light curve of GRB~221009A obtained mainly with Pan-STARRS and a few other telescopes. They assume the X-ray decay index as the GRB afterglow decay index and find excess flux in the $r,i,$ and $y$ filters and not in the $z$ filter. They find the excess flux follows a similar behavior to SN~2016jca and SN~2017iuk. Thus, this excess is explained as an emerging supernova with absolute peak AB magnitudes of $M_g = -19.8 \pm 0.6$ mag, $M_r = -19.5 \pm 0.3$, and $M_z = -20.1 \pm 0.3$. Since we do not have good coverage in the $z$-band light curve and we did not collect any $y$-band, we have not included these bands in our analyses. Whereas, for $r$ and $i$ filters, we have good coverage of the data set and have done a similar analysis. For our data set, we also find excess in flux when compared to a power-law based on the X-ray data. However, our analyses show that this excess is better explained by a simple broken power-law fit to optical data rather than the addition of a bright supernova component with $M_r = -19.5$ mag. The shallowness in the optical decay can be explained by the relation $\alpha_{X} = \alpha_{opt} + 0.25$ expected for the slow cooling regime for constant interstellar medium \citep{Zaninoni_2013}. If we assume the observed data to be explained by a broken power law with a decay index shallower than the XRT data, and also $E(B-V) = 1.74$ mag, in that case, there could be an associated supernova with $M_r = -19.5$ mag which would not show a clear bump in the $r$ band light-curve. 

We note that \citet{Postigo_2022b} and \citet{Rossi_2022} reported via GCN circulars the detection of associated supernova spectroscopically at an average time of 8 and 8.56 days after the BAT trigger. From our light curve analysis, for us to detect supernova bumps during that time period, the absolute magnitude of the associated supernova would have to be brighter than $M_r = -20.66$ after correcting for the extinction of $E(B-V) = 1.32$ mag. This value is greater than the absolute magnitude at the peak for both SN~1998bw ($M_r = -19.41$) and SN~2012bz ($M_r = -19.63$). We also note that our MMT spectra at the last epoch coincide with the predicted peak in the supernova component from the analyses of \citet{Fulton_2023}. However, we do not see any broad features in our spectrum (Fig.~\ref{fig:spectroscopy}) which is indicative of SN type Ic-BL. We note that our spectrum is noisy compared to other SN spectra as shown in Fig.~\ref{fig:spectroscopy_comp} for comparison. 

Alternately, our results could indicate that GRB~221009A has no associated supernova at all. Previously, there have been observations of supernova-less GRBs such as GRB~060505 \citep{Fynbo_2006}, GRB~060614 \citep{Della_2006, Fynbo_2006, Gal_2006}, GRB~111005A \citep{Michal_2018}, and GRB~2211211A \citep{Rastinejad_2022_kilonova}. For all of these cases, unlike GRB~221009A, the source is not located in a region of high extinction. Thus, dust extinction has been ruled out for the three supernova-less GRBs. For the case of GRB~060614, the afterglow is faint and even a supernova similar to SN~2006aj would be clearly visible in the light curve, ruling out the possibility of even a very faint associated supernova. \citet{Dado_2018} suggest that such supernova-less GRBs could comprise half of the GRB population, and could originate from a phase transition of neutron stars to quark stars in high-mass X-ray binaries. 

Recently, the bright GRB~211211A with redshift ($z = 0.0763 \pm 0.0002$) was also found to have no associated supernova \citep{Rastinejad_2022_kilonova, Troja_2022}. Instead, its light curve shows an excess in the NIR which points toward an associated kilonova and a binary neutron star merger origin. In the case of GRB~221009A, we do not detect a strong infrared excess indicative of an associated kilonova and thus consider a compact object merger scenario unlikely for this burst. Long-term monitoring of GRB~221009A, including with powerful facilities like HST and JWST, will improve our understanding of the afterglow model and improve our limits on a possible supernova contribution, providing new tests of these and other models for the progenitors and central engines powering the most energetic GRBs.

\facilities{ADS,  CTIO:DECam, LBT (MODS), MMT (Binospec), HET (LRS2), Las Cumbres Observatory (MUSCAT3),  NED, SOAR (GHTS), WISeREP, HST, IRSA }

\software{ astropy \citep{astropy:2013,astropy:2018}, Photutils \citep{Bradley_2019}, Binospec IDL \citep{BinoIDL}, Panacea, BANZAI \citep{Banzai}, Light Curve Fitting \citep{lightcurvefitting}, MatPLOTLIB \citep{mpl}, NumPy \citep{numpy}, Scipy \citep{scipy}, IRAF \citep{iraf1,iraf2}}



\section*{Acknowledgments}

We are grateful to J. Rastinejad \& W.-F. Fong for help with the MMT observing, and comments on an earlier version of the manuscript.  We also thank B. Weiner for his help with the MMT observations.
We would also like to thank our anonymous referee for constructive comments on the paper which have greatly improved the paper. We also like to thank E. Pian for sharing SN~2013dx and SN~2106jca light curve data.

Time domain research by the University of Arizona team and D.J.S. \ is supported by NSF grants AST-1821987, 1813466, 1908972, \& 2108032, and by the Heising-Simons Foundation under grant \#20201864. 

This publication was made possible through the support of an LSSTC Catalyst Fellowship to K.A.B., funded through Grant 62192 from the John Templeton Foundation to LSST Corporation. The opinions expressed in this publication are those of the authors and do not necessarily reflect the views of LSSTC or the John Templeton Foundation.

Observations reported here were obtained at the MMT Observatory, a joint facility of the University of Arizona and the Smithsonian Institution. This paper uses data taken with the MODS spectrographs built with funding from NSF grant AST-9987045 and the NSF Telescope System Instrumentation Program (TSIP), with additional funds from the Ohio Board of Regents and the Ohio State University Office of Research. 

This research has made use of the NASA Astrophysics Data System (ADS) Bibliographic Services, and the NASA/IPAC Infrared Science Archive (IRSA), which is funded by the National Aeronautics and Space Administration and operated by the California Institute of Technology. This research made use of Photutils, an Astropy package for detection and photometry of astronomical sources (\cite{Bradley_2019}). 

This work made use of data supplied by the UK Swift Science Data Centre at the University of Leicester. The DELVE project is partially supported by the NASA Fermi Guest Investigator Program Cycle 9 No. 91201. This manuscript has been authored by Fermi Research Alliance, LLC under Contract No. DE-AC02-07CH11359 with the U.S. Department of Energy, Office of Science, Office of High Energy Physics. The United States Government retains and the publisher, by accepting the article for publication, acknowledges that the United States Government retains a non-exclusive, paid-up, irrevocable, world wide license to publish or reproduce the published form of this manuscript or allow others to do so, for United States Government purposes. This project used data obtained with the Dark Energy Camera (DECam), which was constructed by the Dark Energy Survey (DES) collaboration. Funding for the DES Projects has been provided by the US Department of Energy, the US National Science Foundation, the Ministry of Science and Education of Spain, the Science and Technology Facilities Council of the United Kingdom, the Higher Education Funding Council for England, the National Center for Supercomputing Applications at the University of Illinois at Urbana-Champaign, the Kavli Institute for Cosmological Physics at the University of Chicago, Center for Cosmology and Astro-Particle Physics at the Ohio State University, the Mitchell Institute for Fundamental Physics and Astronomy at Texas A\&M University, Financiadora de Estudos e Projetos, Fundação Carlos Chagas Filho de Amparo à Pesquisa do Estado do Rio de Janeiro, Conselho Nacional de Desenvolvimento Científico e Tecnológico and the Ministério da Ciência, Tecnologia e Inovação, the Deutsche Forschungsgemeinschaft and the Collaborating Institutions in the Dark Energy Survey.
The Collaborating Institutions are Argonne National Laboratory, the University of California at Santa Cruz, the University of Cambridge, Centro de Investigaciones Enérgeticas, Medioambientales y Tecnológicas–Madrid, the University of Chicago, University College London, the DES-Brazil Consortium, the University of Edinburgh, the Eidgenössische Technische Hochschule (ETH) Zürich, Fermi National Accelerator Laboratory, the University of Illinois at Urbana-Champaign, the Institut de Ciències de l’Espai (IEEC/CSIC), the Institut de Física d’Altes Energies, Lawrence Berkeley National Laboratory, the Ludwig-Maximilians Universität München and the associated Excellence Cluster Universe, the University of Michigan, NSF’s NOIRLab, the University of Nottingham, the Ohio State University, the OzDES Membership Consortium, the University of Pennsylvania, the University of Portsmouth, SLAC National Accelerator Laboratory, Stanford University, the University of Sussex, and Texas A\&M University.

Based on observations at Cerro Tololo Inter-American Observatory, NSF’s NOIRLab (NOIRLab Prop. ID 2022B-827437; PI: G.\ Hosseinzadeh; 2019A-0305; PI: A.\ Drlica-Wagner), which is managed by the Association of Universities for Research in Astronomy (AURA) under a cooperative agreement with the National Science Foundation.

Based in part on observations obtained at the Southern Astrophysical Research (SOAR) telescope, which is a joint project of the Minist\'{e}rio da Ci\^{e}ncia, Tecnologia e Inova\c{c}\~{o}es (MCTI/LNA) do Brasil, the US National Science Foundation’s NOIRLab, the University of North Carolina at Chapel Hill (UNC), and Michigan State University (MSU). The observations reported here were obtained in part at the MMT Observatory, a facility operated jointly by the Smithsonian Institution and the University of Arizona. MMT telescope time was granted by NSF’s NOIRLab, through the Telescope System Instrumentation Program (TSIP). TSIP was funded by NSF (NOIRLab Prop. ID UAO-S143-22B ; PI: S. Wyatt).

This research is based on observations made with the NASA/ESA Hubble Space Telescope obtained from the Space Telescope Science Institute, which is operated by the Association of Universities for Research in Astronomy, Inc., under NASA contract NAS 5–26555. These observations are associated with program 17264. The authors acknowledge the team led by PI A. J. Levan for developing their observing program with a zero-exclusive-access period. 

The Low Resolution Spectrograph 2 (LRS2) was developed and funded by the University of Texas at Austin McDonald Observatory and Department of Astronomy, and by Pennsylvania State University. We thank the Leibniz-Institut fur Astrophysik Potsdam (AIP) and the Institut fur Astrophysik Goettingen (IAG) for their contributions to the construction of the integral field units. Based on observations obtained with the Hobby-Eberly Telescope (HET), which is a joint project of the University of Texas at Austin, the Pennsylvania State University, Ludwig-Maximillians-Universitaet Muenchen, and Georg-August Universitaet Goettingen. The HET is named in honor of its principal benefactors, William P. Hobby and Robert E. Eberly. 

This research has made use of the CfA Supernova Archive, which is funded in part by the National Science Foundation through grant AST 0907903. This work makes use of data taken with the Las Cumbres Observatory global telescope network. The LCO group is supported by NSF grants 1911225 and 1911151.

JEA and CEMV are supported by the international Gemini Observatory, a program of NSF’s NOIRLab, which is managed by the Association of Universities for Research in Astronomy (AURA) under a cooperative agreement with the National Science Foundation, on behalf of the Gemini partnership of Argentina, Brazil, Canada, Chile, the Republic of Korea, and the United States of America. JAC-B acknowledges support from FONDECYT Regular N 1220083.

%
\vspace{5mm}
\bibliography{grb}{}

\newcommand{\noop}[1]{}
\begin{thebibliography}{}
\expandafter\ifx\csname natexlab\endcsname\relax\def\natexlab#1{#1}\fi
\providecommand{\url}[1]{\href{#1}{#1}}
\providecommand{\dodoi}[1]{doi:~\href{http://doi.org/#1}{\nolinkurl{#1}}}
\providecommand{\doeprint}[1]{\href{http://ascl.net/#1}{\nolinkurl{http://ascl.net/#1}}}
\providecommand{\doarXiv}[1]{\href{https://arxiv.org/abs/#1}{\nolinkurl{https://arxiv.org/abs/#1}}}

\bibitem[{{Ashall} {et~al.}(2019){Ashall}, {Mazzali}, {Pian}, {Woosley},
  {Palazzi}, {Prentice}, {Kobayashi}, {Holmbo}, {Levan}, {Perley},
  {Stritzinger}, {Bufano}, {Filippenko}, {Melandri}, {Oates}, {Rossi},
  {Selsing}, {Zheng}, {Castro-Tirado}, {Chincarini}, {D'Avanzo}, {De Pasquale},
  {Emery}, {Fruchter}, {Hurley}, {Moller}, {Nomoto}, {Tanaka}, \&
  {Valeev}}]{Ashall_2019}
{Ashall}, C., {Mazzali}, P.~A., {Pian}, E., {et~al.} 2019, \mnras, 487, 5824,
  \dodoi{10.1093/mnras/stz1588}

\bibitem[{{Astropy Collaboration} {et~al.}(2013){Astropy Collaboration},
  {Robitaille}, {Tollerud}, {Greenfield}, {Droettboom}, {Bray}, {Aldcroft},
  {Davis}, {Ginsburg}, {Price-Whelan}, {Kerzendorf}, {Conley}, {Crighton},
  {Barbary}, {Muna}, {Ferguson}, {Grollier}, {Parikh}, {Nair}, {Unther},
  {Deil}, {Woillez}, {Conseil}, {Kramer}, {Turner}, {Singer}, {Fox}, {Weaver},
  {Zabalza}, {Edwards}, {Azalee Bostroem}, {Burke}, {Casey}, {Crawford},
  {Dencheva}, {Ely}, {Jenness}, {Labrie}, {Lim}, {Pierfederici}, {Pontzen},
  {Ptak}, {Refsdal}, {Servillat}, \& {Streicher}}]{astropy:2013}
{Astropy Collaboration}, {Robitaille}, T.~P., {Tollerud}, E.~J., {et~al.} 2013,
  \aap, 558, A33, \dodoi{10.1051/0004-6361/201322068}

\bibitem[{{Atteia}(2022)}]{Atteia_2022}
{Atteia}, J.~L. 2022, GRB Coordinates Network, 32793, 1

\bibitem[{{Balaji} {et~al.}(2023){Balaji}, {Ramirez-Quezada}, {Silk}, \&
  {Zhang}}]{Balaji_2023}
{Balaji}, S., {Ramirez-Quezada}, M.~E., {Silk}, J., \& {Zhang}, Y. 2023, arXiv
  e-prints, arXiv:2301.02258, \dodoi{10.48550/arXiv.2301.02258}

\bibitem[{Barbary(2016)}]{extinction}
Barbary, K. 2016, extinction v0.3.0,  Zenodo, \dodoi{10.5281/zenodo.804967}

\bibitem[{{Belkin} {et~al.}(2022{\natexlab{a}}){Belkin}, {Kim}, {Pozanenko},
  {Krugov}, {Aimuratov}, {Pankov}, \& {GRB IKI FuN}}]{Belkin_2022b}
{Belkin}, S., {Kim}, V., {Pozanenko}, A., {et~al.} 2022{\natexlab{a}}, GRB
  Coordinates Network, 32769, 1

\bibitem[{{Belkin} {et~al.}(2022{\natexlab{b}}){Belkin}, {Nazarov},
  {Pozanenko}, {Pankov}, \& {IKI GRB FuN}}]{Belkin_2022c}
{Belkin}, S., {Nazarov}, S., {Pozanenko}, A., {Pankov}, N., \& {IKI GRB FuN}.
  2022{\natexlab{b}}, GRB Coordinates Network, 32684, 1

\bibitem[{{Belkin} {et~al.}(2022{\natexlab{c}}){Belkin}, {Pozanenko}, {Klunko},
  {Pankov}, \& {GRB IKI FuN}}]{Belkin_2022}
{Belkin}, S., {Pozanenko}, A., {Klunko}, E., {Pankov}, N., \& {GRB IKI FuN}.
  2022{\natexlab{c}}, GRB Coordinates Network, 32645, 1

\bibitem[{{Bennett} {et~al.}(2014){Bennett}, {Larson}, {Weiland}, \&
  {Hinshaw}}]{Bennett_2014}
{Bennett}, C.~L., {Larson}, D., {Weiland}, J.~L., \& {Hinshaw}, G. 2014, \apj,
  794, 135, \dodoi{10.1088/0004-637X/794/2/135}

\bibitem[{{Bikmaev} {et~al.}(2022{\natexlab{a}}){Bikmaev}, {Khamitov},
  {Irtuganov}, {Gorbachev}, {Sakhibullin}, \& {Burenin}}]{Bikmaev_2022}
{Bikmaev}, I., {Khamitov}, I., {Irtuganov}, E., {et~al.} 2022{\natexlab{a}},
  GRB Coordinates Network, 32743, 1

\bibitem[{{Bikmaev} {et~al.}(2022{\natexlab{b}}){Bikmaev}, {Khamitov},
  {Irtuganov}, {Gorbachev}, {Sakhibullin}, \& {Burenin}}]{Bikmaev_2022b}
---. 2022{\natexlab{b}}, GRB Coordinates Network, 32752, 1

\bibitem[{Bradley {et~al.}(2019)Bradley, Sip{\H o}cz, Robitaille, Tollerud,
  Vin{\'{\i}}cius, Deil, Barbary, G{\"u}nther, Cara, Busko, Conseil,
  Droettboom, Bostroem, Bray, Bratholm, Wilson, Craig, Barentsen, Pascual,
  Donath, Greco, Perren, Lim, \& Kerzendorf}]{Bradley_2019}
Bradley, L., Sip{\H o}cz, B., Robitaille, T., {et~al.} 2019, astropy/photutils:
  v0.6, \dodoi{10.5281/zenodo.2533376}

\bibitem[{{Brivio} {et~al.}(2022){Brivio}, {Ferro}, {D'Avanzo}, {Fugazza},
  {Melandri}, {Covino}, \& {REM Team}}]{Brivio_2022}
{Brivio}, R., {Ferro}, M., {D'Avanzo}, P., {et~al.} 2022, GRB Coordinates
  Network, 32652, 1

\bibitem[{{Brown} {et~al.}(2013){Brown}, {Baliber}, {Bianco}, {Bowman},
  {Burleson}, {Conway}, {Crellin}, {Depagne}, {De Vera}, {Dilday}, {Dragomir},
  {Dubberley}, {Eastman}, {Elphick}, {Falarski}, {Foale}, {Ford}, {Fulton},
  {Garza}, {Gomez}, {Graham}, {Greene}, {Haldeman}, {Hawkins}, {Haworth},
  {Haynes}, {Hidas}, {Hjelstrom}, {Howell}, {Hygelund}, {Lister}, {Lobdill},
  {Martinez}, {Mullins}, {Norbury}, {Parrent}, {Paulson}, {Petry}, {Pickles},
  {Posner}, {Rosing}, {Ross}, {Sand}, {Saunders}, {Shobbrook}, {Shporer},
  {Street}, {Thomas}, {Tsapras}, {Tufts}, {Valenti}, {Vander Horst}, {Walker},
  {White}, \& {Willis}}]{FTN}
{Brown}, T.~M., {Baliber}, N., {Bianco}, F.~B., {et~al.} 2013, \pasp, 125,
  1031, \dodoi{10.1086/673168}

\bibitem[{{Butler} {et~al.}(2022){Butler}, {Watson}, {Dichiara}, {Becerra},
  {Dimitrova}, {L{\'o}pez}, {Gonz{\'a}lez}, {Kutyrev}, {Lee}, {Pereyra}, \&
  {Troja}}]{Butler_2022}
{Butler}, N., {Watson}, A.~M., {Dichiara}, S., {et~al.} 2022, GRB Coordinates
  Network, 32705, 1

\bibitem[{{Cano} {et~al.}(2017{\natexlab{a}}){Cano}, {Wang}, {Dai}, \&
  {Wu}}]{Cano_2017}
{Cano}, Z., {Wang}, S.-Q., {Dai}, Z.-G., \& {Wu}, X.-F. 2017{\natexlab{a}},
  Advances in Astronomy, 2017, 8929054,
  \dodoi{10.1155/2017/892905410.48550/arXiv.1604.03549}

\bibitem[{{Cano} {et~al.}(2017{\natexlab{b}}){Cano}, {Izzo}, {de Ugarte
  Postigo}, {Th{\"o}ne}, {Kr{\"u}hler}, {Heintz}, {Malesani}, {Geier},
  {Fuentes}, {Chen}, {Covino}, {D'Elia}, {Fynbo}, {Goldoni}, {Gomboc},
  {Hjorth}, {Jakobsson}, {Kann}, {Milvang-Jensen}, {Pugliese},
  {S{\'a}nchez-Ram{\'\i}rez}, {Schulze}, {Sollerman}, {Tanvir}, \&
  {Wiersema}}]{Cano_2017b}
{Cano}, Z., {Izzo}, L., {de Ugarte Postigo}, A., {et~al.} 2017{\natexlab{b}},
  \aap, 605, A107, \dodoi{10.1051/0004-6361/201731005}

\bibitem[{{Castro-Tirado} {et~al.}(2022){Castro-Tirado}, {Sanchez-Ramirez},
  {Hu}, {Caballero-Garcia}, {Castro Tirado}, {Fernandez-Garcia},
  {Perez-Garcia}, {Lombardi}, {Pandey}, {Yang}, \& {Zhang}}]{Tirado_2022}
{Castro-Tirado}, A.~J., {Sanchez-Ramirez}, R., {Hu}, Y.~D., {et~al.} 2022, GRB
  Coordinates Network, 32686, 1

\bibitem[{{Chambers} {et~al.}(2016){Chambers}, {Magnier}, {Metcalfe},
  {Flewelling}, {Huber}, {Waters}, {Denneau}, {Draper}, {Farrow}, {Finkbeiner},
  {Holmberg}, {Koppenhoefer}, {Price}, {Rest}, {Saglia}, {Schlafly}, {Smartt},
  {Sweeney}, {Wainscoat}, {Burgett}, {Chastel}, {Grav}, {Heasley}, {Hodapp},
  {Jedicke}, {Kaiser}, {Kudritzki}, {Luppino}, {Lupton}, {Monet}, {Morgan},
  {Onaka}, {Shiao}, {Stubbs}, {Tonry}, {White}, {Ba{\~n}ados}, {Bell},
  {Bender}, {Bernard}, {Boegner}, {Boffi}, {Botticella}, {Calamida},
  {Casertano}, {Chen}, {Chen}, {Cole}, {Deacon}, {Frenk}, {Fitzsimmons},
  {Gezari}, {Gibbs}, {Goessl}, {Goggia}, {Gourgue}, {Goldman}, {Grant},
  {Grebel}, {Hambly}, {Hasinger}, {Heavens}, {Heckman}, {Henderson}, {Henning},
  {Holman}, {Hopp}, {Ip}, {Isani}, {Jackson}, {Keyes}, {Koekemoer}, {Kotak},
  {Le}, {Liska}, {Long}, {Lucey}, {Liu}, {Martin}, {Masci}, {McLean}, {Mindel},
  {Misra}, {Morganson}, {Murphy}, {Obaika}, {Narayan}, {Nieto-Santisteban},
  {Norberg}, {Peacock}, {Pier}, {Postman}, {Primak}, {Rae}, {Rai}, {Riess},
  {Riffeser}, {Rix}, {R{\"o}ser}, {Russel}, {Rutz}, {Schilbach}, {Schultz},
  {Scolnic}, {Strolger}, {Szalay}, {Seitz}, {Small}, {Smith}, {Soderblom},
  {Taylor}, {Thomson}, {Taylor}, {Thakar}, {Thiel}, {Thilker}, {Unger},
  {Urata}, {Valenti}, {Wagner}, {Walder}, {Walter}, {Watters}, {Werner},
  {Wood-Vasey}, \& {Wyse}}]{Chambers_2016}
{Chambers}, K.~C., {Magnier}, E.~A., {Metcalfe}, N., {et~al.} 2016, arXiv
  e-prints, arXiv:1612.05560.
\newblock \doarXiv{1612.05560}

\bibitem[{{Chen} {et~al.}(2022){Chen}, {Malesani}, {Yang}, {Hou}, {Ngeow},
  {Pan}, {Hsiao}, {Lin}, \& {Guo}}]{Chen_2022}
{Chen}, T.~W., {Malesani}, D.~B., {Yang}, S., {et~al.} 2022, GRB Coordinates
  Network, 32667, 1

\bibitem[{{Chonis} {et~al.}(2016){Chonis}, {Hill}, {Lee}, {Tuttle}, {Vattiat},
  {Drory}, {Indahl}, {Peterson}, \& {Ramsey}}]{Chonis16}
{Chonis}, T.~S., {Hill}, G.~J., {Lee}, H., {et~al.} 2016, in Society of
  Photo-Optical Instrumentation Engineers (SPIE) Conference Series, Vol. 9908,
  Ground-based and Airborne Instrumentation for Astronomy VI, ed. C.~J.
  {Evans}, L.~{Simard}, \& H.~{Takami}, 99084C, \dodoi{10.1117/12.2232209}

\bibitem[{{Clemens} {et~al.}(2004){Clemens}, {Crain}, \& {Anderson}}]{soar}
{Clemens}, J.~C., {Crain}, J.~A., \& {Anderson}, R. 2004, in Society of
  Photo-Optical Instrumentation Engineers (SPIE) Conference Series, Vol. 5492,
  Ground-based Instrumentation for Astronomy, ed. A.~F.~M. {Moorwood} \&
  M.~{Iye}, 331--340, \dodoi{10.1117/12.550069}

\bibitem[{{Dado} \& {Dar}(2018)}]{Dado_2018}
{Dado}, S., \& {Dar}, A. 2018, \apj, 855, 88, \dodoi{10.3847/1538-4357/aaad69}

\bibitem[{{Dainotti} {et~al.}(2022){Dainotti}, {De Simone}, {Islam},
  {Kawaguchi}, {Moriya}, {Takiwaki}, {Tominaga}, \&
  {Gangopadhyay}}]{Dainotti_2022}
{Dainotti}, M.~G., {De Simone}, B., {Islam}, K.~M., {et~al.} 2022, \apj, 938,
  41, \dodoi{10.3847/1538-4357/ac8b77}

\bibitem[{{D'Avanzo} {et~al.}(2022){D'Avanzo}, {Ferro}, {Brivio}, {Bernardini},
  {Fugazza}, {Campana}, {Covino}, {D'Elia}, {De Pasquale}, {Malesani},
  {Melandri}, {Palazzi}, {Piranomonte}, {Rossi}, {Sbarufatti}, {Tagliaferri},
  {REM Team}, \& {CIBO Collaboration}}]{Davanzo_2022}
{D'Avanzo}, P., {Ferro}, M., {Brivio}, R., {et~al.} 2022, GRB Coordinates
  Network, 32755, 1

\bibitem[{{de Ugarte Postigo} {et~al.}(2022{\natexlab{a}}){de Ugarte Postigo},
  {Izzo}, {Thoene}, {Fynbo}, {Kann}, {Agui Fernandez}, \&
  {Tanvir}}]{Postigo_2022b}
{de Ugarte Postigo}, A., {Izzo}, L., {Thoene}, C.~C., {et~al.}
  2022{\natexlab{a}}, GRB Coordinates Network, 32800, 1

\bibitem[{{de Ugarte Postigo} {et~al.}(2022{\natexlab{b}}){de Ugarte Postigo},
  {Izzo}, {Pugliese}, {Xu}, {Schneider}, {Fynbo}, {Tanvir}, {Malesani},
  {Saccardi}, {Kann}, {Wiersema}, {Gompertz}, {Thoene}, {Levan}, \& {Stargate
  Collaboration}}]{postigo_2022}
{de Ugarte Postigo}, A., {Izzo}, L., {Pugliese}, G., {et~al.}
  2022{\natexlab{b}}, GRB Coordinates Network, 32648, 1

\bibitem[{{de Wet} {et~al.}(2022){de Wet}, {Groot}, \& {Meerlicht
  Consortium}}]{deWet_2022}
{de Wet}, S., {Groot}, P.~J., \& {Meerlicht Consortium}. 2022, GRB Coordinates
  Network, 32646, 1

\bibitem[{{Della Valle} {et~al.}(2006){Della Valle}, {Chincarini}, {Panagia},
  {Tagliaferri}, {Malesani}, {Testa}, {Fugazza}, {Campana}, {Covino},
  {Mangano}, {Antonelli}, {D'Avanzo}, {Hurley}, {Mirabel}, {Pellizza},
  {Piranomonte}, \& {Stella}}]{Della_2006}
{Della Valle}, M., {Chincarini}, G., {Panagia}, N., {et~al.} 2006, \nat, 444,
  1050, \dodoi{10.1038/nature05374}

\bibitem[{{Dressel}(2022)}]{WFC3_2022}
{Dressel}, L. 2022, in WFC3 Instrument Handbook for Cycle 30 v. 14, Vol.~14
  (STScI), 14

\bibitem[{{Drlica-Wagner} {et~al.}(2021){Drlica-Wagner}, {Carlin}, {Nidever},
  {Ferguson}, {Kuropatkin}, {Adam{\'o}w}, {Cerny}, {Choi}, {Esteves},
  {Mart{\'\i}nez-V{\'a}zquez}, {Mau}, {Miller}, {Mutlu-Pakdil}, {Neilsen},
  {Olsen}, {Pace}, {Riley}, {Sakowska}, {Sand}, {Santana-Silva}, {Tollerud},
  {Tucker}, {Vivas}, {Zaborowski}, {Zenteno}, {Abbott}, {Allam}, {Bechtol},
  {Bell}, {Bell}, {Bilaji}, {Bom}, {Carballo-Bello}, {Crnojevi{\'c}}, {Cioni},
  {Diaz-Ocampo}, {de Boer}, {Erkal}, {Gruendl}, {Hernandez-Lang}, {Hughes},
  {James}, {Johnson}, {Li}, {Mao}, {Mart{\'\i}nez-Delgado}, {Massana},
  {McNanna}, {Morgan}, {Nadler}, {No{\"e}l}, {Palmese}, {Peter}, {Rykoff},
  {S{\'a}nchez}, {Shipp}, {Simon}, {Smercina}, {Soares-Santos}, {Stringfellow},
  {Tavangar}, {van der Marel}, {Walker}, {Wechsler}, {Wu}, {Yanny},
  {Fitzpatrick}, {Huang}, {Jacques}, {Nikutta}, {Scott}, \& {Astro Data
  Lab}}]{Delve1}
{Drlica-Wagner}, A., {Carlin}, J.~L., {Nidever}, D.~L., {et~al.} 2021, \apjs,
  256, 2, \dodoi{10.3847/1538-4365/ac079d}

\bibitem[{{Drlica-Wagner} {et~al.}(2022){Drlica-Wagner}, {Ferguson},
  {Adam{\'o}w}, {Aguena}, {Allam}, {Andrade-Oliveira}, {Bacon}, {Bechtol},
  {Bell}, {Bertin}, {Bilaji}, {Bocquet}, {Bom}, {Brooks}, {Burke},
  {Carballo-Bello}, {Carlin}, {Carnero Rosell}, {Carrasco Kind}, {Carretero},
  {Castander}, {Cerny}, {Chang}, {Choi}, {Conselice}, {Costanzi},
  {Crnojevi{\'c}}, {da Costa}, {de Vicente}, {Desai}, {Esteves}, {Everett},
  {Ferrero}, {Fitzpatrick}, {Flaugher}, {Friedel}, {Frieman},
  {Garc{\'\i}a-Bellido}, {Gatti}, {Gaztanaga}, {Gerdes}, {Gruen}, {Gruendl},
  {Gschwend}, {Hartley}, {Hernandez-Lang}, {Hinton}, {Hollowood}, {Honscheid},
  {Hughes}, {Jacques}, {James}, {Johnson}, {Kuehn}, {Kuropatkin}, {Lahav},
  {Li}, {Lidman}, {Lin}, {March}, {Marshall}, {Mart{\'\i}nez-Delgado},
  {Mart{\'\i}nez-V{\'a}zquez}, {Massana}, {Mau}, {McNanna}, {Melchior},
  {Menanteau}, {Miller}, {Miquel}, {Mohr}, {Morgan}, {Mutlu-Pakdil},
  {Mu{\~n}oz}, {Neilsen}, {Nidever}, {Nikutta}, {Nilo Castellon}, {No{\"e}l},
  {Ogando}, {Olsen}, {Pace}, {Palmese}, {Paz-Chinch{\'o}n}, {Pereira},
  {Pieres}, {Plazas Malag{\'o}n}, {Prat}, {Riley}, {Rodriguez-Monroy}, {Romer},
  {Roodman}, {Sako}, {Sakowska}, {Sanchez}, {S{\'a}nchez}, {Sand},
  {Santana-Silva}, {Santiago}, {Schubnell}, {Serrano}, {Sevilla-Noarbe},
  {Simon}, {Smith}, {Soares-Santos}, {Stringfellow}, {Suchyta}, {Suson}, {Tan},
  {Tarle}, {Tavangar}, {Thomas}, {To}, {Tollerud}, {Troxel}, {Tucker}, {Varga},
  {Vivas}, {Walker}, {Weller}, {Wilkinson}, {Wu}, {Yanny}, {Zaborowski},
  {Zenteno}, {Delve Collaboration}, {Des Collaboration}, \& {Astro Data
  Lab}}]{Delve2}
{Drlica-Wagner}, A., {Ferguson}, P.~S., {Adam{\'o}w}, M., {et~al.} 2022, \apjs,
  261, 38, \dodoi{10.3847/1538-4365/ac78eb}

\bibitem[{{Drout} {et~al.}(2011){Drout}, {Soderberg}, {Gal-Yam}, {Cenko},
  {Fox}, {Leonard}, {Sand}, {Moon}, {Arcavi}, \& {Green}}]{Drout_2011}
{Drout}, M.~R., {Soderberg}, A.~M., {Gal-Yam}, A., {et~al.} 2011, \apj, 741,
  97, \dodoi{10.1088/0004-637X/741/2/9710.48550/arXiv.1011.4959}

\bibitem[{{Dzhappuev} {et~al.}(2022){Dzhappuev}, {Afashokov}, {Dzaparova},
  {Dzhatdoev}, {Gorbacheva}, {Karpikov}, {Khadzhiev}, {Klimenko}, {Kudzhaev},
  {Kurenya}, {Lidvansky}, {Mikhailova}, {Petkov}, {Podlesnyi}, {Pozdnukhov},
  {Romanenko}, {Rubtsov}, {Troitsky}, {Unatlokov}, {Vaiman}, {Yanin}, \&
  {Zhuravleva}}]{Dzhappuev_2022}
{Dzhappuev}, D.~D., {Afashokov}, Y.~Z., {Dzaparova}, I.~M., {et~al.} 2022, The
  Astronomer's Telegram, 15669, 1

\bibitem[{{Evans} {et~al.}(2009){Evans}, {Beardmore}, {Page}, {Osborne},
  {O'Brien}, {Willingale}, {Starling}, {Burrows}, {Godet}, {Vetere}, {Racusin},
  {Goad}, {Wiersema}, {Angelini}, {Capalbi}, {Chincarini}, {Gehrels}, {Kennea},
  {Margutti}, {Morris}, {Mountford}, {Pagani}, {Perri}, {Romano}, \&
  {Tanvir}}]{Evans_2009}
{Evans}, P.~A., {Beardmore}, A.~P., {Page}, K.~L., {et~al.} 2009, \mnras, 397,
  1177, \dodoi{10.1111/j.1365-2966.2009.14913.x}

\bibitem[{{Fabricant} {et~al.}(2019{\natexlab{a}}){Fabricant}, {Fata}, {Epps},
  {Gauron}, {Mueller}, {Zajac}, {Amato}, {Barberis}, {Bergner}, {Brennan},
  {Brown}, {Chilingarian}, {Geary}, {Kradinov}, {McLeod}, {Smith}, \&
  {Woods}}]{Binospec}
{Fabricant}, D., {Fata}, R., {Epps}, H., {et~al.} 2019{\natexlab{a}}, \pasp,
  131, 075004, \dodoi{10.1088/1538-3873/ab1d78}

\bibitem[{{Fabricant} {et~al.}(2019{\natexlab{b}}){Fabricant}, {Fata}, {Epps},
  {Gauron}, {Mueller}, {Zajac}, {Amato}, {Barberis}, {Bergner}, {Brennan},
  {Brown}, {Chilingarian}, {Geary}, {Kradinov}, {McLeod}, {Smith}, \&
  {Woods}}]{Fabrican_2019}
---. 2019{\natexlab{b}}, \pasp, 131, 075004, \dodoi{10.1088/1538-3873/ab1d78}

\bibitem[{{Ferro} {et~al.}(2022){Ferro}, {Brivio}, {D'Avanzo}, {Piranomonte},
  {Lorenzi}, {Mainella}, \& {CIBO Collaboration}}]{Ferro_2022}
{Ferro}, M., {Brivio}, R., {D'Avanzo}, P., {et~al.} 2022, GRB Coordinates
  Network, 32804, 1

\bibitem[{{Filippenko}(1997)}]{Filippenko_1997}
{Filippenko}, A.~V. 1997, \araa, 35, 309,
  \dodoi{10.1146/annurev.astro.35.1.309}

\bibitem[{Flaugher {et~al.}(2015)Flaugher, Diehl, Honscheid, Abbott, Alvarez,
  Angstadt, Annis, Antonik, Ballester, Beaufore, Bernstein, Bernstein, Bigelow,
  Bonati, Boprie, Brooks, Buckley-Geer, Campa, Cardiel-Sas, Castander,
  Castilla, Cease, Cela-Ruiz, Chappa, Chi, Cooper, da~Costa, Dede, Derylo,
  DePoy, de~Vicente, Doel, Drlica-Wagner, Eiting, Elliott, Emes, Estrada, Neto,
  Finley, Flores, Frieman, Gerdes, Gladders, Gregory, Gutierrez, Hao, Holland,
  Holm, Huffman, Jackson, James, Jonas, Karcher, Karliner, Kent, Kessler,
  Kozlovsky, Kron, Kubik, Kuehn, Kuhlmann, Kuk, Lahav, Lathrop, Lee, Levi,
  Lewis, Li, Mandrichenko, Marshall, Martinez, Merritt, Miquel, Muñoz,
  Neilsen, Nichol, Nord, Ogando, Olsen, Palaio, Patton, Peoples, Plazas, Rauch,
  Reil, Rheault, Roe, Rogers, Roodman, Sanchez, Scarpine, Schindler, Schmidt,
  Schmitt, Schubnell, Schultz, Schurter, Scott, Serrano, Shaw, Smith,
  Soares-Santos, Stefanik, Stuermer, Suchyta, Sypniewski, Tarle, Thaler, Tighe,
  Tran, Tucker, Walker, Wang, Watson, Weaverdyck, Wester, Woods, Yanny, \&
  Collaboration)}]{Flaugher_2015}
Flaugher, B., Diehl, H.~T., Honscheid, K., {et~al.} 2015, The Astronomical
  Journal, 150, 150, \dodoi{10.1088/0004-6256/150/5/150}

\bibitem[{{Fulton} {et~al.}(2023){Fulton}, {Smartt}, {Rhodes}, {Huber},
  {Villar}, {Moore}, {Srivastav}, {Schultz}, {Chambers}, {Izzo}, {Hjorth},
  {Chen}, {Nicholl}, {Foley}, {Rest}, {Smith}, {Young}, {Sim}, {Bright},
  {Zenati}, {de Boer}, {Bulger}, {Fairlamb}, {Gao}, {Lin}, {Lowe}, {Magnier},
  {Smith}, {Wainscoat}, {Coulter}, {Jones}, {Kilpatrick}, {McGill},
  {Ramirez-Ruiz}, {Lee}, {Narayan}, {Ramakrishnan}, {Ridden-Harper}, {Singh},
  {Wang}, {Kong}, {Ngeow}, {Pan}, {Yang}, {Davis}, {Piro}, {Rojas-Bravo},
  {Sommer}, \& {Yadavalli}}]{Fulton_2023}
{Fulton}, M.~D., {Smartt}, S.~J., {Rhodes}, L., {et~al.} 2023, arXiv e-prints,
  arXiv:2301.11170.
\newblock \doarXiv{2301.11170}

\bibitem[{{Fynbo} {et~al.}(2006){Fynbo}, {Watson}, {Th{\"o}ne}, {Sollerman},
  {Bloom}, {Davis}, {Hjorth}, {Jakobsson}, {J{\o}rgensen}, {Graham},
  {Fruchter}, {Bersier}, {Kewley}, {Cassan}, {Castro Cer{\'o}n}, {Foley},
  {Gorosabel}, {Hinse}, {Horne}, {Jensen}, {Klose}, {Kocevski}, {Marquette},
  {Perley}, {Ramirez-Ruiz}, {Stritzinger}, {Vreeswijk}, {Wijers}, {Woller},
  {Xu}, \& {Zub}}]{Fynbo_2006}
{Fynbo}, J. P.~U., {Watson}, D., {Th{\"o}ne}, C.~C., {et~al.} 2006, \nat, 444,
  1047, \dodoi{10.1038/nature05375}

\bibitem[{{Gal-Yam} {et~al.}(2006){Gal-Yam}, {Fox}, {Price}, {Ofek}, {Davis},
  {Leonard}, {Soderberg}, {Schmidt}, {Lewis}, {Peterson}, {Kulkarni}, {Berger},
  {Cenko}, {Sari}, {Sharon}, {Frail}, {Moon}, {Brown}, {Cucchiara}, {Harrison},
  {Piran}, {Persson}, {McCarthy}, {Penprase}, {Chevalier}, \&
  {MacFadyen}}]{Gal_2006}
{Gal-Yam}, A., {Fox}, D.~B., {Price}, P.~A., {et~al.} 2006, \nat, 444, 1053,
  \dodoi{10.1038/nature05373}

\bibitem[{{Galama} {et~al.}(1998){Galama}, {Vreeswijk}, {van Paradijs},
  {Kouveliotou}, {Augusteijn}, {B{\"o}hnhardt}, {Brewer}, {Doublier},
  {Gonzalez}, {Leibundgut}, {Lidman}, {Hainaut}, {Patat}, {Heise}, {in't Zand},
  {Hurley}, {Groot}, {Strom}, {Mazzali}, {Iwamoto}, {Nomoto}, {Umeda},
  {Nakamura}, {Young}, {Suzuki}, {Shigeyama}, {Koshut}, {Kippen}, {Robinson},
  {de Wildt}, {Wijers}, {Tanvir}, {Greiner}, {Pian}, {Palazzi}, {Frontera},
  {Masetti}, {Nicastro}, {Feroci}, {Costa}, {Piro}, {Peterson}, {Tinney},
  {Boyle}, {Cannon}, {Stathakis}, {Sadler}, {Begam}, \& {Ianna}}]{Galama_1998}
{Galama}, T.~J., {Vreeswijk}, P.~M., {van Paradijs}, J., {et~al.} 1998, \nat,
  395, 670, \dodoi{10.1038/27150}

\bibitem[{{Gaskell} {et~al.}(1986){Gaskell}, {Cappellaro}, {Dinerstein},
  {Garnett}, {Harkness}, \& {Wheeler}}]{Gaskell_1986}
{Gaskell}, C.~M., {Cappellaro}, E., {Dinerstein}, H.~L., {et~al.} 1986, \apjl,
  306, L77, \dodoi{10.1086/184709}

\bibitem[{{Groot} {et~al.}(2022){Groot}, {Vreeswijk}, {Ter Horst}, {Bloemen},
  {Jonker}, {de Wet}, {Malesani}, {Pieterse}, \& {BlackGEM
  Consortium}}]{Groot_2022}
{Groot}, P.~J., {Vreeswijk}, P.~M., {Ter Horst}, R., {et~al.} 2022, GRB
  Coordinates Network, 32678, 1

\bibitem[{Harris {et~al.}(2020)Harris, Millman, van~der Walt, Gommers,
  Virtanen, Cournapeau, Wieser, Taylor, Berg, Smith, Kern, Picus, Hoyer, van
  Kerkwijk, Brett, Haldane, del R{\'{i}}o, Wiebe, Peterson,
  G{\'{e}}rard-Marchant, Sheppard, Reddy, Weckesser, Abbasi, Gohlke, \&
  Oliphant}]{numpy}
Harris, C.~R., Millman, K.~J., van~der Walt, S.~J., {et~al.} 2020, Nature, 585,
  357, \dodoi{10.1038/s41586-020-2649-2}

\bibitem[{{Hill} {et~al.}(2021){Hill}, {Lee}, {MacQueen}, {Kelz}, {Drory},
  {Vattiat}, {Good}, {Ramsey}, {Kriel}, {Peterson}, {DePoy}, {Gebhardt},
  {Marshall}, {Tuttle}, {Bauer}, {Chonis}, {Fabricius}, {Froning},
  {H{\"a}user}, {Indahl}, {Jahn}, {Landriau}, {Leck}, {Montesano}, {Prochaska},
  {Snigula}, {Zeimann}, {Bryant}, {Damm}, {Fowler}, {Janowiecki}, {Martin},
  {Mrozinski}, {Odewahn}, {Rostopchin}, {Shetrone}, {Spencer}, {Mentuch
  Cooper}, {Armandroff}, {Bender}, {Dalton}, {Hopp}, {Komatsu}, {Nicklas},
  {Ramsey}, {Roth}, {Schneider}, {Sneden}, \& {Steinmetz}}]{het2}
{Hill}, G.~J., {Lee}, H., {MacQueen}, P.~J., {et~al.} 2021, \aj, 162, 298,
  \dodoi{10.3847/1538-3881/ac2c02}

\bibitem[{Hjorth \& Bloom(2012)}]{Hjorth_2012}
Hjorth, J., \& Bloom, J.~S. 2012, The Gamma-Ray Burst - Supernova Connection
  (Cambridge University Press), 169--190

\bibitem[{{Hosseinzadeh} \& {Gomez}(2020)}]{lightcurvefitting}
{Hosseinzadeh}, G., \& {Gomez}, S. 2020, {Light Curve Fitting}, v0.2.0, Zenodo,
   Zenodo, \dodoi{10.5281/zenodo.4312178}

\bibitem[{{Hosseinzadeh} \& {Gomez}(2022)}]{hosseinzadeh_2022}
---. 2022, Light Curve Fitting, v0.7.0,  Zenodo, \dodoi{10.5281/zenodo.7250571}

\bibitem[{{Hu} {et~al.}(2022){Hu}, {Casanova}, {Fernandez-Garcia}, {Castro
  Tirado}, {Caballero-Garcia}, {Olivares}, {Perez-Garcia}, {Sanchez-Ramirez},
  {Castro-Tirado}, {Perez del Pulgar}, {Castellon}, {Fernandez-Munoz}, \&
  {Jelinek}}]{Hu_2022}
{Hu}, Y.~D., {Casanova}, V., {Fernandez-Garcia}, E., {et~al.} 2022, GRB
  Coordinates Network, 32644, 1

\bibitem[{{Huang} {et~al.}(2022){Huang}, {Hu}, {Chen}, {Zha}, {Liu}, {Yao},
  {Cao}, \& {Experiment}}]{Huang_2022}
{Huang}, Y., {Hu}, S., {Chen}, S., {et~al.} 2022, GRB Coordinates Network,
  32677, 1

\bibitem[{{Huber} {et~al.}(2022){Huber}, {Schultz}, {Chambers}, {Smith},
  {Fulton}, {Smartt}, {Chen}, {Nicholl}, {Young}, {Shingles}, {Srivastav},
  {Sim}, {de Boer}, {Bulger}, {Fairlamb}, {Lin}, {Lowe}, {Magnier},
  {Wainscoat}, {Gao}, {Stubbs}, \& {Rest}}]{Huber_2022}
{Huber}, M., {Schultz}, A., {Chambers}, K.~C., {et~al.} 2022, GRB Coordinates
  Network, 32758, 1

\bibitem[{{Hunter}(2007)}]{mpl}
{Hunter}, J.~D. 2007, Computing in Science and Engineering, 9, 90,
  \dodoi{10.1109/MCSE.2007.55}

\bibitem[{{Iwamoto} {et~al.}(1998){Iwamoto}, {Mazzali}, {Nomoto}, {Umeda},
  {Nakamura}, {Patat}, {Danziger}, {Young}, {Suzuki}, {Shigeyama},
  {Augusteijn}, {Doublier}, {Gonzalez}, {Boehnhardt}, {Brewer}, {Hainaut},
  {Lidman}, {Leibundgut}, {Cappellaro}, {Turatto}, {Galama}, {Vreeswijk},
  {Kouveliotou}, {van Paradijs}, {Pian}, {Palazzi}, \&
  {Frontera}}]{Iwamoto_1998}
{Iwamoto}, K., {Mazzali}, P.~A., {Nomoto}, K., {et~al.} 1998, \nat, 395, 672,
  \dodoi{10.1038/27155}

\bibitem[{{Izzo} {et~al.}(2022){Izzo}, {Saccardi}, {Fynbo}, {Palmerio},
  {Malesani}, {Agui Fernandez}, {Kann}, {Melandri}, {Vergani}, {Wiersema}, \&
  {Stargate Consortium}}]{Izzo_2022}
{Izzo}, L., {Saccardi}, A., {Fynbo}, J.~P.~U., {et~al.} 2022, GRB Coordinates
  Network, 32765, 1

\bibitem[{{Kann} {et~al.}(2023){Kann}, {Agayeva}, {Aivazyan}, {Alishov},
  {Andrade}, {Antier}, {Baransky}, {Bendjoya}, {Benkhaldoun}, {Beradze},
  {Berezin}, {Bo{\"e}r}, {Broens}, {Brunier}, {Bulla}, {Burkhonov}, {Burns},
  {Chen}, {Chen}, {Conti}, {Coughlin}, {Cui}, {Daigne}, {Delaveau},
  {Devillepoix}, {Dietrich}, {Dornic}, {Dubois}, {Ducoin}, {Durand}, {Duverne},
  {Eggenstein}, {Ehgamberdiev}, {Fouad}, {Freeberg}, {Froebrich}, {Ge},
  {Gervasoni}, {Godunova}, {Gokuldass}, {Gurbanov}, {Han}, {Hasanov}, {Hello},
  {Hussenot-Desenonges}, {Inasaridze}, {Iskandar}, {Ismailov}, {Janati}, {Jegou
  du Laz}, {Jia}, {Karpov}, {Kaeouach}, {Kiendrebeogo}, {Klotz}, {Kneip},
  {Kochiashvili}, {Kunert}, {Lekic}, {Leonini}, {Li}, {Li}, {Li}, {Liao},
  {Logie}, {Lu}, {Mao}, {Marchais}, {M{\'e}nard}, {Morris}, {Natsvlishvili},
  {Nedora}, {Noonan}, {Noysena}, {Orange}, {Pang}, {Peng}, {Pellouin},
  {Peloton}, {Pradier}, {Pyshna}, {Rajabo}, {Rau}, {Rinner}, {Rivet},
  {Romanov}, {Rosi}, {Rupchandani}, {Serrau}, {Shokry}, {Simon}, {Smith},
  {Sokoliuk}, {Soliman}, {Song}, {Takey}, {Tillayev}, {Tinjaca Ramirez},
  {Melo}, {Turpin}, {de Ugarte Postigo}, {Vanaverbeke}, {Vasylenko}, {Vernet},
  {Vidadi}, {Wang}, {Wang}, {Wang}, {Wang}, {Xiong}, {Xu}, {Xue}, {Zeng},
  {Zhang}, {Zhao}, \& {Zhao}}]{Kann_2023}
{Kann}, D.~A., {Agayeva}, S., {Aivazyan}, V., {et~al.} 2023, arXiv e-prints,
  arXiv:2302.06225.
\newblock \doarXiv{2302.06225}

\bibitem[{{Kansky} {et~al.}(2019){Kansky}, {Chilingarian}, {Fabricant},
  {Matthews}, {Moran}, {Paegert}, {Duane Gibson}, {Porter}, \&
  {Roll}}]{BinoIDL}
{Kansky}, J., {Chilingarian}, I., {Fabricant}, D., {et~al.} 2019, \pasp, 131,
  075005, \dodoi{10.1088/1538-3873/ab1ceb}

\bibitem[{Kass \& Raftery(1995)}]{Kass_1995}
Kass, R.~E., \& Raftery, A.~E. 1995, Journal of the American Statistical
  Association, 90, 773, \dodoi{10.1080/01621459.1995.10476572}

\bibitem[{{Kim} {et~al.}(2022){Kim}, {Krugov}, {Pozanenko}, {Aimuratov},
  {Belkin}, {Pankov}, \& {IKI FuN}}]{Kim_2022}
{Kim}, V., {Krugov}, M., {Pozanenko}, A., {et~al.} 2022, GRB Coordinates
  Network, 32670, 1

\bibitem[{{Kimura} {et~al.}(2022){Kimura}, {Isogai}, {Arimoto}, {Yonetoku},
  {Narita}, {Tamura}, {Fukui}, \& {Ikoma}}]{Kimura_2022}
{Kimura}, M., {Isogai}, K., {Arimoto}, M., {et~al.} 2022, GRB Coordinates
  Network, 33038, 1

\bibitem[{{Kulkarni} {et~al.}(1998){Kulkarni}, {Frail}, {Wieringa}, {Ekers},
  {Sadler}, {Wark}, {Higdon}, {Phinney}, \& {Bloom}}]{Kulkarni_1998}
{Kulkarni}, S.~R., {Frail}, D.~A., {Wieringa}, M.~H., {et~al.} 1998, \nat, 395,
  663, \dodoi{10.1038/27139}

\bibitem[{{Kumar} {et~al.}(2022){Kumar}, {Swain}, {Waratkar}, {Angail},
  {Bhalerao}, {Anupama}, {Barway}, \& {GIT Team}}]{Kumar_2022}
{Kumar}, H., {Swain}, V., {Waratkar}, G., {et~al.} 2022, GRB Coordinates
  Network, 32662, 1

\bibitem[{{Laskar} {et~al.}(2022){Laskar}, {Alexander}, {Ayache}, {Berger},
  {Chornock}, {van Eerten}, {Fong}, {Margutti}, {Mundell}, \&
  {Schady}}]{Laskar_2022}
{Laskar}, T., {Alexander}, K.~D., {Ayache}, E., {et~al.} 2022, GRB Coordinates
  Network, 32757, 1

\bibitem[{{Laskar} {et~al.}(\noop{2023}submitted){Laskar}, {Alexander},
  {Margutti}, {Tarraneh}, {Chornock}, {Berger}, {Cendes}, {Duerr}, {Perley},
  {Ravasio}, {Yamazaki}, {Ayache}, {Barclay}, {Duran}, {Bhandari}, {Brethauer},
  {Christy}, {Coppejans}, {Duffell}, {Fong}, {Gomboc}, {Guidorzi}, {Kennea},
  {Kobayashi}, {Levan}, {Lobanov}, {Metzger}, {Ros}, {Schroeder}, \&
  {Williams}}]{Laskar_2023}
{Laskar}, T., {Alexander}, K.~D., {Margutti}, R., {et~al.}
  \noop{2023}submitted, \apjl

\bibitem[{{Levan} {et~al.}(2022){Levan}, {Barclay}, {Bhirombhakdi}, {Burns},
  {Cenko}, {Chrimes}, {D'Avanzo}, {D'Elia}, {Della Valle}, {de Ugarte Postigo},
  {Fong}, {Fruchter}, {Gompertz}, {Hartmann}, {Hedges}, {Heintz}, {Izzo},
  {Jakobsson}, {Jonker}, {Kann}, {Kennea}, {Le Floc'h}, {Malesani}, {Melandri},
  {Metzger}, {Mullally}, {Pian}, {Piranomonte}, {Pugliese}, {Racusin},
  {Rastinejad}, {Ravasio}, {Rossi}, {Salvaterra}, {Sbarufatti}, {Schneider},
  {Starling}, {Tanvir}, {Thoene}, {Vergani}, {Wijers}, \& {Xu}}]{Levan_2022}
{Levan}, A.~J., {Barclay}, T., {Bhirombhakdi}, K., {et~al.} 2022, GRB
  Coordinates Network, 32921, 1

\bibitem[{{Levan} {et~al.}(2023){Levan}, {Lamb}, {Schneider}, {Hjorth},
  {Zafar}, {de Ugarte Postigo}, {Sargent}, {Mullally}, {Izzo}, {D'Avanzo},
  {Burns}, {Ag{\"u}{\'\i} Fern{\'a}ndez}, {Barclay}, {Bernardini},
  {Bhirombhakdi}, {Bremer}, {Brivio}, {Campana}, {Chrimes}, {D'Elia}, {De
  Pasquale}, {Ferro}, {Fong}, {Fruchter}, {Fynbo}, {Gaspari}, {Gompertz},
  {Hartmann}, {Hedges}, {Heintz}, {Hotokezaka}, {Jakobsson}, {Kann}, {Kennea},
  {Laskar}, {Le Floc'h}, {Malesani}, {Melandri}, {Metzger}, {Oates}, {Pian},
  {Piranomonte}, {Pugliese}, {Racusin}, {Rastinejad}, {Ravasio}, {Rossi},
  {Saccardi}, {Salvaterra}, {Sbarufatti}, {Starling}, {Tanvir}, {Th{\"o}ne},
  {Vergani}, {Watson}, {Wiersema}, \& {Xu}}]{Levan_2023}
{Levan}, A.~J., {Lamb}, G.~P., {Schneider}, B., {et~al.} 2023, arXiv e-prints,
  arXiv:2302.07761.
\newblock \doarXiv{2302.07761}

\bibitem[{{MAGIC Collaboration} {et~al.}(2019){MAGIC Collaboration}, {Acciari},
  {Ansoldi}, {Antonelli}, {Arbet Engels}, {Baack}, {Babi{\'c}}, {Banerjee},
  {Barres de Almeida}, {Barrio}, {Becerra Gonz{\'a}lez}, {Bednarek},
  {Bellizzi}, {Bernardini}, {Berti}, {Besenrieder}, {Bhattacharyya},
  {Bigongiari}, {Biland}, {Blanch}, {Bonnoli}, {Bo{\v{s}}njak}, {Busetto},
  {Carosi}, {Carosi}, {Ceribella}, {Chai}, {Chilingaryan}, {Cikota}, {Colak},
  {Colin}, {Colombo}, {Contreras}, {Cortina}, {Covino}, {D'Amico}, {D'Elia},
  {da Vela}, {Dazzi}, {de Angelis}, {de Lotto}, {Delfino}, {Delgado},
  {Depaoli}, {di Pierro}, {di Venere}, {Do Souto Espi{\~n}eira}, {Dominis
  Prester}, {Donini}, {Dorner}, {Doro}, {Elsaesser}, {Fallah Ramazani},
  {Fattorini}, {Fern{\'a}ndez-Barral}, {Ferrara}, {Fidalgo}, {Foffano},
  {Fonseca}, {Font}, {Fruck}, {Fukami}, {Gallozzi}, {Garc{\'\i}a L{\'o}pez},
  {Garczarczyk}, {Gasparyan}, {Gaug}, {Giglietto}, {Giordano}, {Godinovi{\'c}},
  {Green}, {Guberman}, {Hadasch}, {Hahn}, {Herrera}, {Hoang}, {Hrupec},
  {H{\"u}tten}, {Inada}, {Inoue}, {Ishio}, {Iwamura}, {Jouvin}, {Kerszberg},
  {Kubo}, {Kushida}, {Lamastra}, {Lelas}, {Leone}, {Lindfors}, {Lombardi},
  {Longo}, {L{\'o}pez}, {L{\'o}pez-Coto}, {L{\'o}pez-Oramas}, {Loporchio},
  {Machado de Oliveira Fraga}, {Maggio}, {Majumdar}, {Makariev}, {Mallamaci},
  {Maneva}, {Manganaro}, {Mannheim}, {Maraschi}, {Mariotti}, {Mart{\'\i}nez},
  {Masuda}, {Mazin}, {Mi{\'c}anovi{\'c}}, {Miceli}, {Minev}, {Miranda},
  {Mirzoyan}, {Molina}, {Moralejo}, {Morcuende}, {Moreno}, {Moretti},
  {Munar-Adrover}, {Neustroev}, {Nigro}, {Nilsson}, {Ninci}, {Nishijima},
  {Noda}, {Nogu{\'e}s}, {N{\"o}the}, {Nozaki}, {Paiano}, {Palacio},
  {Palatiello}, {Paneque}, {Paoletti}, {Paredes}, {Pe{\~n}il}, {Peresano},
  {Persic}, {Prada Moroni}, {Prandini}, {Puljak}, {Rhode}, {Rib{\'o}}, {Rico},
  {Righi}, {Rugliancich}, {Saha}, {Sahakyan}, {Saito}, {Sakurai}, {Satalecka},
  {Schmidt}, {Schweizer}, {Sitarek}, {{\v{S}}nidari{\'c}}, {Sobczynska},
  {Somero}, {Stamerra}, {Strom}, {Strzys}, {Suda}, {Suri{\'c}}, {Takahashi},
  {Tavecchio}, {Temnikov}, {Terzi{\'c}}, {Teshima}, {Torres-Alb{\`a}}, {Tosti},
  {Tsujimoto}, {Vagelli}, {van Scherpenberg}, {Vanzo}, {Vazquez Acosta},
  {Vigorito}, {Vitale}, {Vovk}, {Will}, {Zari{\'c}}, \& {Nava}}]{Magic_2019}
{MAGIC Collaboration}, {Acciari}, V.~A., {Ansoldi}, S., {et~al.} 2019, \nat,
  575, 455, \dodoi{10.1038/s41586-019-1750-x}

\bibitem[{{Mao} {et~al.}(2022){Mao}, {Lu}, {Zhao}, \& {Bai}}]{Mao_2022}
{Mao}, J., {Lu}, K.~X., {Zhao}, X.~H., \& {Bai}, J.~M. 2022, GRB Coordinates
  Network, 32727, 1

\bibitem[{{Mazzali} {et~al.}(2021){Mazzali}, {Pian}, {Bufano}, \&
  {Ashall}}]{Mazzali_2021}
{Mazzali}, P.~A., {Pian}, E., {Bufano}, F., \& {Ashall}, C. 2021, \mnras, 505,
  4106, \dodoi{10.1093/mnras/stab1594}

\bibitem[{{McCully} {et~al.}(2018){McCully}, {Volgenau}, {Harbeck}, {Lister},
  {Saunders}, {Turner}, {Siiverd}, \& {Bowman}}]{Banzai}
{McCully}, C., {Volgenau}, N.~H., {Harbeck}, D.-R., {et~al.} 2018, in Society
  of Photo-Optical Instrumentation Engineers (SPIE) Conference Series, Vol.
  10707, Software and Cyberinfrastructure for Astronomy V, ed. J.~C. {Guzman}
  \& J.~{Ibsen}, 107070K, \dodoi{10.1117/12.2314340}

\bibitem[{{Melandri} {et~al.}(2022){Melandri}, {Izzo}, {Pian}, {Malesani},
  {Della Valle}, {Rossi}, {D'Avanzo}, {Guetta}, {Mazzali}, {Benetti},
  {Masetti}, {Palazzi}, {Savaglio}, {Amati}, {Antonelli}, {Ashall},
  {Bernardini}, {Campana}, {Carini}, {Covino}, {D'Elia}, {de Ugarte Postigo},
  {De Pasquale}, {Filippenko}, {Fruchter}, {Fynbo}, {Giunta}, {Hartmann},
  {Jakobsson}, {Japelj}, {Jonker}, {Kann}, {Lamb}, {Levan}, {Martin-Carrillo},
  {M{\o}ller}, {Piranomonte}, {Pugliese}, {Salvaterra}, {Schulze}, {Starling},
  {Stella}, {Tagliaferri}, {Tanvir}, \& {Watson}}]{Melandri_2022}
{Melandri}, A., {Izzo}, L., {Pian}, E., {et~al.} 2022, \aap, 659, A39,
  \dodoi{10.1051/0004-6361/202141788}

\bibitem[{{Micha{\l}owski} {et~al.}(2018){Micha{\l}owski}, {Xu}, {Stevens},
  {Levan}, {Yang}, {Paragi}, {Kamble}, {Tsai}, {Dannerbauer}, {van der Horst},
  {Shao}, {Crosby}, {Gentile}, {Stanway}, {Wiersema}, {Fynbo}, {Tanvir},
  {Kamphuis}, {Garrett}, \& {Bartczak}}]{Michal_2018}
{Micha{\l}owski}, M.~J., {Xu}, D., {Stevens}, J., {et~al.} 2018, \aap, 616,
  A169, \dodoi{10.1051/0004-6361/201629942}

\bibitem[{{Mirabal}(2023)}]{Mirabal_2023}
{Mirabal}, N. 2023, \mnras, 519, L85, \dodoi{10.1093/mnrasl/slac157}

\bibitem[{{Modjaz} {et~al.}(2011){Modjaz}, {Kewley}, {Bloom}, {Filippenko},
  {Perley}, \& {Silverman}}]{Modjaz_2011}
{Modjaz}, M., {Kewley}, L., {Bloom}, J.~S., {et~al.} 2011, \apjl, 731, L4,
  \dodoi{10.1088/2041-8205/731/1/L410.48550/arXiv.1007.0661}

\bibitem[{{Modjaz} {et~al.}(2016){Modjaz}, {Liu}, {Bianco}, \&
  {Graur}}]{Modjaz_2016}
{Modjaz}, M., {Liu}, Y.~Q., {Bianco}, F.~B., \& {Graur}, O. 2016, \apj, 832,
  108, \dodoi{10.3847/0004-637X/832/2/10810.48550/arXiv.1509.07124}

\bibitem[{{Modjaz} {et~al.}(2006){Modjaz}, {Stanek}, {Garnavich}, {Berlind},
  {Blondin}, {Brown}, {Calkins}, {Challis}, {Diamond-Stanic}, {Hao}, {Hicken},
  {Kirshner}, \& {Prieto}}]{Modjaz_2006}
{Modjaz}, M., {Stanek}, K.~Z., {Garnavich}, P.~M., {et~al.} 2006, \apjl, 645,
  L21, \dodoi{10.1086/505906}

\bibitem[{{Modjaz} {et~al.}(2020){Modjaz}, {Bianco}, {Siwek}, {Huang},
  {Perley}, {Fierroz}, {Liu}, {Arcavi}, {Gal-Yam}, {Filippenko},
  {Blagorodnova}, {Cenko}, {Kasliwal}, {Kulkarni}, {Schulze}, {Taggart}, \&
  {Zheng}}]{Modjaz_2020}
{Modjaz}, M., {Bianco}, F.~B., {Siwek}, M., {et~al.} 2020, \apj, 892, 153,
  \dodoi{10.3847/1538-4357/ab418510.48550/arXiv.1901.00872}

\bibitem[{{Narita} {et~al.}(2020){Narita}, {Fukui}, {Yamamuro}, {Harbeck},
  {Bowman}, {Elphick}, {Nation}, {Armstrong}, {Han}, {Abe}, {Ikoma}, {Isogai},
  {Kawauchi}, {Kurita}, {Kusakabe}, {de Leon}, {Livingston}, {Mori},
  {Nishiumi}, {Tamura}, {Watanabe}, {Volgenau}, {Heinrich-Josties}, {Foale},
  {Daily}, {McCully}, {Kirby}, {Smith}, {Haworth}, {Conway},
  {Storrie-Lombardi}, {Rosing}, {Chatelain}, {Bachelet}, {Johnson}, \&
  {Rabus}}]{muscat}
{Narita}, N., {Fukui}, A., {Yamamuro}, T., {et~al.} 2020, in Society of
  Photo-Optical Instrumentation Engineers (SPIE) Conference Series, Vol. 11447,
  Society of Photo-Optical Instrumentation Engineers (SPIE) Conference Series,
  114475K, \dodoi{10.1117/12.2559947}

\bibitem[{{Nomoto} {et~al.}(1995){Nomoto}, {Iwamoto}, \&
  {Suzuki}}]{Nomoto_1995}
{Nomoto}, K.~I., {Iwamoto}, K., \& {Suzuki}, T. 1995, \physrep, 256, 173,
  \dodoi{10.1016/0370-1573(94)00107-E}

\bibitem[{{O'Connor} {et~al.}(2022{\natexlab{a}}){O'Connor}, {Cenko}, {Troja},
  {Dichiara}, {Kutyrev}, {Veilleux}, \& {Durbak}}]{Connor_2022b}
{O'Connor}, B., {Cenko}, S.~B., {Troja}, E., {et~al.} 2022{\natexlab{a}}, GRB
  Coordinates Network, 32739, 1

\bibitem[{{O'Connor} {et~al.}(2022{\natexlab{b}}){O'Connor}, {Cenko}, {Troja},
  {Dichiara}, {Kutyrev}, {Veilleux}, \& {Durbak}}]{Connor_2022}
---. 2022{\natexlab{b}}, GRB Coordinates Network, 32799, 1

\bibitem[{{O'Connor} {et~al.}(2022{\natexlab{c}}){O'Connor}, {Troja},
  {Dichiara}, {Gillanders}, \& {Cenko}}]{oconnor_2022}
{O'Connor}, B., {Troja}, E., {Dichiara}, S., {Gillanders}, J., \& {Cenko},
  S.~B. 2022{\natexlab{c}}, GRB Coordinates Network, 32750, 1

\bibitem[{{O'Connor} {et~al.}(2022{\natexlab{d}}){O'Connor}, {Troja},
  {Dichiara}, {Gillanders}, \& {Cenko}}]{oconnor_2022b}
---. 2022{\natexlab{d}}, GRB Coordinates Network, 32860, 1

\bibitem[{{Paek} {et~al.}(2022){Paek}, {Im}, {Urata}, \& {Sung}}]{Paek_2022}
{Paek}, G. S.~H., {Im}, M., {Urata}, Y., \& {Sung}, H.-I. 2022, GRB Coordinates
  Network, 32659, 1

\bibitem[{{Patat} {et~al.}(2001){Patat}, {Cappellaro}, {Danziger}, {Mazzali},
  {Sollerman}, {Augusteijn}, {Brewer}, {Doublier}, {Gonzalez}, {Hainaut},
  {Lidman}, {Leibundgut}, {Nomoto}, {Nakamura}, {Spyromilio}, {Rizzi},
  {Turatto}, {Walsh}, {Galama}, {van Paradijs}, {Kouveliotou}, {Vreeswijk},
  {Frontera}, {Masetti}, {Palazzi}, \& {Pian}}]{Patat_2001}
{Patat}, F., {Cappellaro}, E., {Danziger}, J., {et~al.} 2001, \apj, 555, 900,
  \dodoi{10.1086/321526}

\bibitem[{{Perley}(2022)}]{Perley_2022}
{Perley}, D.~A. 2022, GRB Coordinates Network, 32638, 1

\bibitem[{{Pian} {et~al.}(2006){Pian}, {Mazzali}, {Masetti}, {Ferrero},
  {Klose}, {Palazzi}, {Ramirez-Ruiz}, {Woosley}, {Kouveliotou}, {Deng},
  {Filippenko}, {Foley}, {Fynbo}, {Kann}, {Li}, {Hjorth}, {Nomoto}, {Patat},
  {Sauer}, {Sollerman}, {Vreeswijk}, {Guenther}, {Levan}, {O'Brien}, {Tanvir},
  {Wijers}, {Dumas}, {Hainaut}, {Wong}, {Baade}, {Wang}, {Amati}, {Cappellaro},
  {Castro-Tirado}, {Ellison}, {Frontera}, {Fruchter}, {Greiner}, {Kawabata},
  {Ledoux}, {Maeda}, {M{\o}ller}, {Nicastro}, {Rol}, \& {Starling}}]{Pian_2006}
{Pian}, E., {Mazzali}, P.~A., {Masetti}, N., {et~al.} 2006, \nat, 442, 1011,
  \dodoi{10.1038/nature05082}

\bibitem[{{Podsiadlowski} {et~al.}(1993){Podsiadlowski}, {Hsu}, {Joss}, \&
  {Ross}}]{Podsiadlowski_1993}
{Podsiadlowski}, P., {Hsu}, J.~J.~L., {Joss}, P.~C., \& {Ross}, R.~R. 1993,
  \nat, 364, 509, \dodoi{10.1038/364509a0}

\bibitem[{{Pogge} {et~al.}(2010){Pogge}, {Atwood}, {Brewer}, {Byard},
  {Derwent}, {Gonzalez}, {Martini}, {Mason}, {O'Brien}, {Osmer}, {Pappalardo},
  {Steinbrecher}, {Teiga}, \& {Zhelem}}]{MODS}
{Pogge}, R.~W., {Atwood}, B., {Brewer}, D.~F., {et~al.} 2010, in Society of
  Photo-Optical Instrumentation Engineers (SPIE) Conference Series, Vol. 7735,
  Ground-based and Airborne Instrumentation for Astronomy III, ed. I.~S.
  {McLean}, S.~K. {Ramsay}, \& H.~{Takami}, 77350A, \dodoi{10.1117/12.857215}

\bibitem[{{Popowski} {et~al.}(2003){Popowski}, {Cook}, \&
  {Becker}}]{Popowski_2003}
{Popowski}, P., {Cook}, K.~H., \& {Becker}, A.~C. 2003, \aj, 126, 2910,
  \dodoi{10.1086/379291}

\bibitem[{{Price-Whelan} {et~al.}(2018){Price-Whelan}, {Sip{\H{o}}cz},
  {G{\"u}nther}, {Lim}, {Crawford}, {Conseil}, {Shupe}, {Craig}, {Dencheva},
  {Ginsburg}, {VanderPlas}, {Bradley}, {P{\'e}rez-Su{\'a}rez}, {de Val-Borro},
  {Paper Contributors}, {Aldcroft}, {Cruz}, {Robitaille}, {Tollerud},
  {Coordination Committee}, {Ardelean}, {Babej}, {Bach}, {Bachetti}, {Bakanov},
  {Bamford}, {Barentsen}, {Barmby}, {Baumbach}, {Berry}, {Biscani}, {Boquien},
  {Bostroem}, {Bouma}, {Brammer}, {Bray}, {Breytenbach}, {Buddelmeijer},
  {Burke}, {Calderone}, {Cano Rodr{\'\i}guez}, {Cara}, {Cardoso}, {Cheedella},
  {Copin}, {Corrales}, {Crichton}, {D{\textquoteright}Avella}, {Deil},
  {Depagne}, {Dietrich}, {Donath}, {Droettboom}, {Earl}, {Erben}, {Fabbro},
  {Ferreira}, {Finethy}, {Fox}, {Garrison}, {Gibbons}, {Goldstein}, {Gommers},
  {Greco}, {Greenfield}, {Groener}, {Grollier}, {Hagen}, {Hirst}, {Homeier},
  {Horton}, {Hosseinzadeh}, {Hu}, {Hunkeler}, {Ivezi{\'c}}, {Jain}, {Jenness},
  {Kanarek}, {Kendrew}, {Kern}, {Kerzendorf}, {Khvalko}, {King}, {Kirkby},
  {Kulkarni}, {Kumar}, {Lee}, {Lenz}, {Littlefair}, {Ma}, {Macleod},
  {Mastropietro}, {McCully}, {Montagnac}, {Morris}, {Mueller}, {Mumford},
  {Muna}, {Murphy}, {Nelson}, {Nguyen}, {Ninan}, {N{\"o}the}, {Ogaz}, {Oh},
  {Parejko}, {Parley}, {Pascual}, {Patil}, {Patil}, {Plunkett}, {Prochaska},
  {Rastogi}, {Reddy Janga}, {Sabater}, {Sakurikar}, {Seifert}, {Sherbert},
  {Sherwood-Taylor}, {Shih}, {Sick}, {Silbiger}, {Singanamalla}, {Singer},
  {Sladen}, {Sooley}, {Sornarajah}, {Streicher}, {Teuben}, {Thomas},
  {Tremblay}, {Turner}, {Terr{\'o}n}, {van Kerkwijk}, {de la Vega}, {Watkins},
  {Weaver}, {Whitmore}, {Woillez}, {Zabalza}, \& {Contributors}}]{astropy:2018}
{Price-Whelan}, A.~M., {Sip{\H{o}}cz}, B.~M., {G{\"u}nther}, H.~M., {et~al.}
  2018, \aj, 156, 123, \dodoi{10.3847/1538-3881/aabc4f}

\bibitem[{{Rajabov} {et~al.}(2022){Rajabov}, {Sadibekova}, {Tillayev},
  {Rinner}, {Benkhaldoun}, {Wang}, {Zhu}, {Zeng}, {Wang}, {Iskandar}, {Fouad},
  {Shokry}, {Takey}, {Soliman}, {Hello}, {Hussenot}, {Boer}, {de Ugarte
  Postigo}, {Antier}, {Kann}, {Burns}, {Simon}, {Baransky}, {Abe}, {Bendjoya},
  {Rivet}, {Vernet}, {Brunier}, {Inasaridze}, {Natsvlishvili}, {Kochiashvili},
  {Beradze}, {Aivazyan}, {Kapanadze}, {Burkhonov}, {Ducoin}, {Ehgamberdiev},
  {Klotz}, {Tosta E. Melo}, \& {GRANDMA Collaboration}}]{Rajabov_2022}
{Rajabov}, Y., {Sadibekova}, T., {Tillayev}, Y., {et~al.} 2022, GRB Coordinates
  Network, 32795, 1

\bibitem[{{Ramsey} {et~al.}(1998){Ramsey}, {Adams}, {Barnes}, {Booth},
  {Cornell}, {Fowler}, {Gaffney}, {Glaspey}, {Good}, {Hill}, {Kelton},
  {Krabbendam}, {Long}, {MacQueen}, {Ray}, {Ricklefs}, {Sage}, {Sebring},
  {Spiesman}, \& {Steiner}}]{het1}
{Ramsey}, L.~W., {Adams}, M.~T., {Barnes}, T.~G., {et~al.} 1998, in Society of
  Photo-Optical Instrumentation Engineers (SPIE) Conference Series, Vol. 3352,
  Advanced Technology Optical/IR Telescopes VI, ed. L.~M. {Stepp}, 34--42,
  \dodoi{10.1117/12.319287}

\bibitem[{{Rastinejad} \& {Fong}(2022)}]{Rastinejad_2022}
{Rastinejad}, J., \& {Fong}, W. 2022, GRB Coordinates Network, 32749, 1

\bibitem[{{Rastinejad} {et~al.}(2022){Rastinejad}, {Gompertz}, {Levan}, {Fong},
  {Nicholl}, {Lamb}, {Malesani}, {Nugent}, {Oates}, {Tanvir}, {de Ugarte
  Postigo}, {Kilpatrick}, {Moore}, {Metzger}, {Ravasio}, {Rossi}, {Schroeder},
  {Jencson}, {Sand}, {Smith}, {Ag{\"u}{\'\i} Fern{\'a}ndez}, {Berger},
  {Blanchard}, {Chornock}, {Cobb}, {De Pasquale}, {Fynbo}, {Izzo}, {Kann},
  {Laskar}, {Marini}, {Paterson}, {Escorial}, {Sears}, \&
  {Th{\"o}ne}}]{Rastinejad_2022_kilonova}
{Rastinejad}, J.~C., {Gompertz}, B.~P., {Levan}, A.~J., {et~al.} 2022, \nat,
  612, 223, \dodoi{10.1038/s41586-022-05390-w}

\bibitem[{{Romanov}(2022{\natexlab{a}})}]{Romanov_2022b}
{Romanov}, F.~D. 2022{\natexlab{a}}, GRB Coordinates Network, 32664, 1

\bibitem[{{Romanov}(2022{\natexlab{b}})}]{Romanov_2022}
---. 2022{\natexlab{b}}, GRB Coordinates Network, 32679, 1

\bibitem[{{Rossi} {et~al.}(2022{\natexlab{a}}){Rossi}, {Maiorano}, {Malesani},
  {CIBO Collaboration}, {Cusano}, \& {Paris}}]{Rossi_2022}
{Rossi}, A., {Maiorano}, E., {Malesani}, D.~B., {et~al.} 2022{\natexlab{a}},
  GRB Coordinates Network, 32809, 1

\bibitem[{{Rossi} {et~al.}(2022{\natexlab{b}}){Rossi}, {Rothberg}, {Palazzi},
  {Kann}, {D'Avanzo}, {Amati}, {Klose}, {Perego}, {Pian}, {Guidorzi},
  {Pozanenko}, {Savaglio}, {Stratta}, {Agapito}, {Covino}, {Cusano}, {D'Elia},
  {Pasquale}, {Della Valle}, {Kuhn}, {Izzo}, {Loffredo}, {Masetti}, {Melandri},
  {Minaev}, {Guelbenzu}, {Paris}, {Paiano}, {Plantet}, {Rossi}, {Salvaterra},
  {Schulze}, {Veillet}, \& {Volnova}}]{Rossi_2022b}
{Rossi}, A., {Rothberg}, B., {Palazzi}, E., {et~al.} 2022{\natexlab{b}}, \apj,
  932, 1, \dodoi{10.3847/1538-4357/ac60a2}

\bibitem[{{Rowles} \& {Froebrich}(2009)}]{Rowles_2009}
{Rowles}, J., \& {Froebrich}, D. 2009, \mnras, 395, 1640,
  \dodoi{10.1111/j.1365-2966.2009.14655.x}

\bibitem[{{Sasada} {et~al.}(2022){Sasada}, {Imai}, {Murata}, {Hosokawa},
  {Niwano}, {Takahashi}, {Tateda}, {Ito}, {Sato}, {Higuchi}, {Yatsu}, {Kawai},
  \& {MITSuME Collaboration}}]{Sasada_2022}
{Sasada}, M., {Imai}, Y., {Murata}, K.~L., {et~al.} 2022, GRB Coordinates
  Network, 32730, 1

\bibitem[{{Schlafly} \& {Finkbeiner}(2011)}]{Schlafly_2011}
{Schlafly}, E.~F., \& {Finkbeiner}, D.~P. 2011, \apj, 737, 103,
  \dodoi{10.1088/0004-637X/737/2/103}

\bibitem[{{Schlegel} {et~al.}(1998){Schlegel}, {Finkbeiner}, \&
  {Davis}}]{Schlegel_1998}
{Schlegel}, D.~J., {Finkbeiner}, D.~P., \& {Davis}, M. 1998, \apj, 500, 525,
  \dodoi{10.1086/305772}

\bibitem[{{Schneider} {et~al.}(2022){Schneider}, {Adami}, {Le Floc'h},
  {Turpin}, {G{\"o}tz}, {Vergani}, {Saccardi}, {Basa}, {Le Van Suu}, \& {a
  larger Collaboration}}]{Schneider_2022}
{Schneider}, B., {Adami}, C., {Le Floc'h}, E., {et~al.} 2022, GRB Coordinates
  Network, 32753, 1

\bibitem[{{Shetrone} {et~al.}(2007){Shetrone}, {Cornell}, {Fowler}, {Gaffney},
  {Laws}, {Mader}, {Mason}, {Odewahn}, {Roman}, {Rostopchin}, {Schneider},
  {Umbarger}, \& {Westfall}}]{het3}
{Shetrone}, M., {Cornell}, M.~E., {Fowler}, J.~R., {et~al.} 2007, \pasp, 119,
  556, \dodoi{10.1086/519291}

\bibitem[{{Shresta} {et~al.}(2022){Shresta}, {Sand}, {Alexander}, {Andrews},
  {Pearson}, {Smith}, \& {Bostroem}}]{Shrestha_2022a}
{Shresta}, M., {Sand}, D., {Alexander}, K.~D., {et~al.} 2022, GRB Coordinates
  Network, 32759, 1

\bibitem[{{Shrestha} {et~al.}(2022){Shrestha}, {Bostroem}, {Sand}, {Alexander},
  {Andrews}, {Pearson}, {Hosseinzadeh}, {Smith}, {Howell}, {McCully},
  {Newsome}, {Padilla Gonzalez}, {Pellegrino}, {Terreran}, {Farah}, \& {Global
  Supernova Project Collaboration}}]{Shrestha_2022b}
{Shrestha}, M., {Bostroem}, K., {Sand}, D., {et~al.} 2022, GRB Coordinates
  Network, 32771, 1

\bibitem[{{Smartt}(2009)}]{Smartt_2009}
{Smartt}, S.~J. 2009, \araa, 47, 63,
  \dodoi{10.1146/annurev-astro-082708-10173710.48550/arXiv.0908.0700}

\bibitem[{{Strausbaugh} \& {Cucchiara}(2022{\natexlab{a}})}]{Strausbaugh_2022}
{Strausbaugh}, R., \& {Cucchiara}, A. 2022{\natexlab{a}}, GRB Coordinates
  Network, 32693, 1

\bibitem[{{Strausbaugh} \& {Cucchiara}(2022{\natexlab{b}})}]{Strausbaugh_2022b}
---. 2022{\natexlab{b}}, GRB Coordinates Network, 32738, 1

\bibitem[{{Tanga} {et~al.}(2018){Tanga}, {Kr{\"u}hler}, {Schady}, {Klose},
  {Graham}, {Greiner}, {Kann}, \& {Nardini}}]{Tanga_2018}
{Tanga}, M., {Kr{\"u}hler}, T., {Schady}, P., {et~al.} 2018, \aap, 615, A136,
  \dodoi{10.1051/0004-6361/201731799}

\bibitem[{{Taubenberger} {et~al.}(2006){Taubenberger}, {Pastorello}, {Mazzali},
  {Valenti}, {Pignata}, {Sauer}, {Arbey}, {B{\"a}rnbantner}, {Benetti}, {Della
  Valle}, {Deng}, {Elias-Rosa}, {Filippenko}, {Foley}, {Goobar}, {Kotak}, {Li},
  {Meikle}, {Mendez}, {Patat}, {Pian}, {Ries}, {Ruiz-Lapuente}, {Salvo},
  {Stanishev}, {Turatto}, \& {Hillebrandt}}]{Taubenberger_2006}
{Taubenberger}, S., {Pastorello}, A., {Mazzali}, P.~A., {et~al.} 2006, \mnras,
  371, 1459, \dodoi{10.1111/j.1365-2966.2006.10776.x}

\bibitem[{{Tody}(1986)}]{iraf1}
{Tody}, D. 1986, in Society of Photo-Optical Instrumentation Engineers (SPIE)
  Conference Series, Vol. 627, Instrumentation in astronomy VI, ed. D.~L.
  {Crawford}, 733, \dodoi{10.1117/12.968154}

\bibitem[{{Tody}(1993)}]{iraf2}
{Tody}, D. 1993, in Astronomical Society of the Pacific Conference Series,
  Vol.~52, Astronomical Data Analysis Software and Systems II, ed. R.~J.
  {Hanisch}, R.~J.~V. {Brissenden}, \& J.~{Barnes}, 173

\bibitem[{{Tomita} {et~al.}(2006){Tomita}, {Deng}, {Maeda}, {Yoshii}, {Nomoto},
  {Mazzali}, {Suzuki}, {Kobayashi}, {Minezaki}, {Aoki}, {Enya}, \&
  {Suganuma}}]{Tomita_2006}
{Tomita}, H., {Deng}, J., {Maeda}, K., {et~al.} 2006, \apj, 644, 400,
  \dodoi{10.1086/503554}

\bibitem[{{Toy} {et~al.}(2016){Toy}, {Cenko}, {Silverman}, {Butler},
  {Cucchiara}, {Watson}, {Bersier}, {Perley}, {Margutti}, {Bellm}, {Bloom},
  {Cao}, {Capone}, {Clubb}, {Corsi}, {De Cia}, {de Diego}, {Filippenko}, {Fox},
  {Gal-Yam}, {Gehrels}, {Georgiev}, {Gonz{\'a}lez}, {Kasliwal}, {Kelly},
  {Kulkarni}, {Kutyrev}, {Lee}, {Prochaska}, {Ramirez-Ruiz}, {Richer},
  {Rom{\'a}n-Z{\'u}{\~n}iga}, {Singer}, {Stern}, {Troja}, \&
  {Veilleux}}]{Toy_2016}
{Toy}, V.~L., {Cenko}, S.~B., {Silverman}, J.~M., {et~al.} 2016, \apj, 818, 79,
  \dodoi{10.3847/0004-637X/818/1/79}

\bibitem[{{Troja} {et~al.}(2022){Troja}, {Fryer}, {O'Connor}, {Ryan},
  {Dichiara}, {Kumar}, {Ito}, {Gupta}, {Wollaeger}, {Norris}, {Kawai},
  {Butler}, {Aryan}, {Misra}, {Hosokawa}, {Murata}, {Niwano}, {Pandey},
  {Kutyrev}, {van Eerten}, {Chase}, {Hu}, {Caballero-Garcia}, \&
  {Castro-Tirado}}]{Troja_2022}
{Troja}, E., {Fryer}, C.~L., {O'Connor}, B., {et~al.} 2022, \nat, 612, 228,
  \dodoi{10.1038/s41586-022-05327-3}

\bibitem[{{Valdes} {et~al.}(2014){Valdes}, {Gruendl}, \& {DES
  Project}}]{decam_pipeline}
{Valdes}, F., {Gruendl}, R., \& {DES Project}. 2014, in Astronomical Society of
  the Pacific Conference Series, Vol. 485, Astronomical Data Analysis Software
  and Systems XXIII, ed. N.~{Manset} \& P.~{Forshay}, 379

\bibitem[{{Veres} {et~al.}(2022){Veres}, {Burns}, {Bissaldi}, {Lesage},
  {Roberts}, \& {Fermi GBM Team}}]{Veres_2022}
{Veres}, P., {Burns}, E., {Bissaldi}, E., {et~al.} 2022, GRB Coordinates
  Network, 32636, 1

\bibitem[{{Vidal} {et~al.}(2022){Vidal}, {Zheng}, {Filippenko}, \& {KAIT GRB
  team}}]{Vidal_2022}
{Vidal}, E., {Zheng}, W., {Filippenko}, A.~V., \& {KAIT GRB team}. 2022, GRB
  Coordinates Network, 32669, 1

\bibitem[{{Vinko} {et~al.}(2022){Vinko}, {Bodi}, {Pal}, {Kriskovics},
  {Szakats}, \& {Vida}}]{Vinko_2022}
{Vinko}, J., {Bodi}, A., {Pal}, A., {et~al.} 2022, GRB Coordinates Network,
  32709, 1

\bibitem[{{Virtanen} {et~al.}(2020){Virtanen}, {Gommers}, {Oliphant},
  {Haberland}, {Reddy}, {Cournapeau}, {Burovski}, {Peterson}, {Weckesser},
  {Bright}, {van der Walt}, {Brett}, {Wilson}, {Millman}, {Mayorov}, {Nelson},
  {Jones}, {Kern}, {Larson}, {Carey}, {Polat}, {Feng}, {Moore}, {VanderPlas},
  {Laxalde}, {Perktold}, {Cimrman}, {Henriksen}, {Quintero}, {Harris},
  {Archibald}, {Ribeiro}, {Pedregosa}, {van Mulbregt}, \& {SciPy 1. 0
  Contributors}}]{scipy}
{Virtanen}, P., {Gommers}, R., {Oliphant}, T.~E., {et~al.} 2020, Nature
  Methods, 17, 261, \dodoi{10.1038/s41592-019-0686-2}

\bibitem[{{Woosley} \& {Bloom}(2006)}]{Woosley_2006}
{Woosley}, S.~E., \& {Bloom}, J.~S. 2006, \araa, 44, 507,
  \dodoi{10.1146/annurev.astro.43.072103.15055810.48550/arXiv.astro-ph/0609142}

\bibitem[{{Xu} {et~al.}(2022){Xu}, {Jiang}, {Fu}, {Liu}, {Zhu}, {Lu}, {Gao}, \&
  {Liu}}]{Xu_2022}
{Xu}, D., {Jiang}, S.~Q., {Fu}, S.~Y., {et~al.} 2022, GRB Coordinates Network,
  32647, 1

\bibitem[{{Yang} {et~al.}(2020){Yang}, {Hoeflich}, {Baade}, {Maund}, {Wang},
  {Brown}, {Stevance}, {Arcavi}, {Burke}, {Cikota}, {Clocchiatti}, {Gal-Yam},
  {Graham}, {Hiramatsu}, {Hosseinzadeh}, {Howell}, {Jha}, {McCully}, {Patat},
  {Sand}, {Schulze}, {Spyromilio}, {Valenti}, {Vink{\'o}}, {Wang}, {Wheeler},
  {Yaron}, \& {Zhang}}]{Yang20}
{Yang}, Y., {Hoeflich}, P., {Baade}, D., {et~al.} 2020, \apj, 902, 46,
  \dodoi{10.3847/1538-4357/aba759}

\bibitem[{{Yaron} \& {Gal-Yam}(2012)}]{Yaron_2012}
{Yaron}, O., \& {Gal-Yam}, A. 2012, \pasp, 124, 668, \dodoi{10.1086/666656}

\bibitem[{{Zaninoni} {et~al.}(2013){Zaninoni}, {Bernardini}, {Margutti},
  {Oates}, \& {Chincarini}}]{Zaninoni_2013}
{Zaninoni}, E., {Bernardini}, M.~G., {Margutti}, R., {Oates}, S., \&
  {Chincarini}, G. 2013, \aap, 557, A12, \dodoi{10.1051/0004-6361/201321221}

\bibitem[{{Zaznobin} {et~al.}(2022){Zaznobin}, {Burenin}, \&
  {Eselevich}}]{Zaznobin_2022}
{Zaznobin}, I., {Burenin}, R., \& {Eselevich}, M. 2022, GRB Coordinates
  Network, 32729, 1

\end{thebibliography}
\bibliographystyle{aasjournal}
\end{document}